\documentclass[12pt]{article}
\usepackage{a4}
\begin{document}
%%%%%%%%%%%%%%%%%%%%%%%%%%%%%%%%%%%%%%%%%%%%%%%%%%%%%%%%%%%%%%%%%%%%%%%%
% several bibliographies in one text file
% the series of new commands, tested on LaTeX2e and on Latex2.09
% via the compatibility mode.
% version redefining thebibliography and cite.
% J. Bijnens 18 Sep. 1998
%%%%%%%%%%%%%%%%%%%%%%%%%%%%%%%%%%%%%%%%%%%%%%%%%%%%%%%%%%%%%%%%%%%%%%%%
\newcounter{zyxabstract}     %  step by one when going to new abstract
\newcounter{zyxrefers}        %  counts the references in every bibliography

\newcommand{\newabstract}
{\newpage\stepcounter{zyxabstract}\setcounter{equation}{0}
\setcounter{footnote}{0}}

\newcommand{\rlabel}[1]{\label{zyx\arabic{zyxabstract}#1}}
\newcommand{\rref}[1]{\ref{zyx\arabic{zyxabstract}#1}}

\renewenvironment{thebibliography}[1] % as standard,the width of largest number
{\section*{References}\setcounter{zyxrefers}{0}
\begin{list}{ [\arabic{zyxrefers}]}{\usecounter{zyxrefers}}}
{\end{list}}

\renewcommand{\bibitem}[1]{\item\rlabel{y#1}}% extra y to avoid duplication
\renewcommand{\cite}[1]{[\rref{y#1}]}      %of labels in text and bibliography
%%%%%%%%%%%%%%%%%%%%%%%%%%%%%%%%%%%%%%%%%%%%%%%%%%%%%%%%%%%%%%%%%%%%%%%%%%%%
%\special{papersize=21cm,28.6cm}
%%%%%%%%%%%
%included: 
%ecker
%talavera
%golterman
%pallante
%neufeld
%guesken
%vladikas
%donoghue
%prades
%kim
%dambrosio
%knecht
%moussallam
%peris
%derafael
%rusetsky
%borasoy
%kambor
%hemmert
%mojzis
%fettes
%hoehler
%steininger
%sainio
%epelbaoum
%savage
%nkaiser
%beane
%leutwyler
%becher
%rkaiser
%diakonov
%stern
%colangelo
%gasser
%wanders
%schechter
%pich
%nyffeler
%goity
%girlanda
%orellana
%toublan
%%%%%%%%%%%
\begin{titlepage}
\begin{flushright}
\small{
FZJ-IKP(TH)-1999-01\\
LU-TP 99-1
}
\end{flushright}

\vspace{1cm}

\begin{center}
{\huge\bf CHIRAL EFFECTIVE\\[1cm] THEORIES}\\[1cm]
205. WE-Heraeus-Seminar\\
Physikzentrum Bad Honnef, Bad Honnef, Germany\\
November 30 --- December 4, 1998\\[2cm]
{\bf Johan Bijnens}\\[0.3cm]
Department of Theoretical Physics 2, Lund University\\
S\"olvegatan 14A, S22362 Lund, Sweden\\[1cm]
{\bf Ulf-G. Mei\ss ner}\\[0.3cm]
{FZ J\"ulich, IKP(Th), D-52425 J\"ulich, Germany}\\[2cm]
{\large ABSTRACT}
\end{center}
These are the proceedings of the workshop on ``Chiral Effective Theories''
held at the Physikzentrum Bad Honnef of the
Deutsche Physikalische Gesellschaft, Bad Honnef, Germany from November 30
to December 4, 1998. The workshop concentrated on Chiral Perturbation
Theory in its various settings and its relations with lattice QCD
and dispersion theory. Included are a short contribution per talk
and a listing of some review papers on the subject.

\end{titlepage}

\section{Introduction}
The field of Chiral Perturbation Theory is a growing area in theoretical
physics. We therefore decided to organize the next topical workshop.
This meeting followed the series of workshops in Ringberg (Germany), 1988,
Dobog\'ok\"o (Hungary), 1991, Karreb\ae ksminde (Denmark), 1993
and Trento (Italy), 1996. All these workshops shared the same features,
about 50 participants, a fairly large amount of time devoted to discussions
rather than presentations and an intimate environment with lots of
discussion opportunities.

This meeting took place in late fall 1998 in the Physikzentrum Bad Honnef
in Bad Honnef, Germany and the funding provided by the Dr. Wilhelm Heinrich
Heraeus und Else Heraeus--Stiftung allowed us to provide for the local
expenses for all participants and to support the travel of some
participants. The WE-Heraeus foundation also provided
the administrative support for the workshop in the person of the
able secretary Jutta Lang. We extend our sincere gratitude to the WE-Heraeus
Stiftung for this support. We would also like to thank the staff of the
Physikzentrum for the excellent service given us during the workshop
and last but not least the participants for making this an exciting
and lively meeting.

The meeting had 53 participants whose names, institutes and email addresses
are listed below. 43 of them presented results in presentations of various
lengths.
A short description of their contents and a list of the most relevant
references can be found below. As in the previous two of these workshops we
felt that this was more appropriate a framework than full-fledged proceedings.
Most results are or will soon be published and available on the archives
so this way we can achieve speedy publication and avoid duplication of
results in the archives.

Below follows first the program, then the list of participants and
a subjective list of review papers, lectures and other proceedings
relevant to the subject of this workshop.
\\[1cm]
\hfill Johan Bijnens and Ulf-G. Mei\ss ner

\newpage
\section{The Program}
\begin{tabbing}
xx.xx \= a very long name\= \kill

{\bf Monday}\\*
12.30 \> Lunch\\
14.00 \> U.-G. Mei\ss ner\\\> E. Dreisigacker \> Opening and Welcome\\
14.20 \>  G.~Ecker \>Introduction to CHPT\\
14.35 \>  G.~Ecker \>The generating functional
at next-to-next-to-leading\\ \>\> order in CHPT\\
15.25 \>  P.~Talavera \> Pion form factors
at $p^6$\\
15.50 \> Coffee\\
16.20 \>   M.~Golterman \>  Quenched and Partially
Quenched Chiral Logarithms\\
17.00 \>  E. Pallante \> The Generating Functional for
Hadronic Weak\\ \>\> Interactions and its Quenched Approximation\\
17.40 \>  H. Neufeld \>A Super-Heat-Kernel Representation for
the One-Loop\\ \>\> Functional of Boson-Fermion systems\\
18.10 \> End of Session\\
18.30 \> Dinner followed by a social evening\\[1cm]

{\bf Tuesday}\\*
9.00 \>  S. G\"usken \>Some Low-Energy
constants from lattice QCD\\
9.40 \>   A. Vladikas \> Quark masses and the
chiral condensate\\ \> \> from the lattice\\
10.10 \>   J. Donoghue \>Dispersive Calculation of
Weak Matrix Elements\\
10.55 \> Coffee\\
11.25 \> J. Prades \>$\Delta S=1$ Transitions
in the $1/N_c$ Expansion\\
12.05 \>  H.-C. Kim \> Effective $\Delta S=1$ Chiral Lagrangian
in the Chiral\\ \>\> Quark Model\\
12.35 \>  G. D'Ambrosio \>Vector Meson Dominance
in the Delta S=1\\ \> \> Weak Chiral Lagrangian\\
13.00\> End of Session\\
14.20 \>  M.~Knecht \>Electromagnetic Corrections to
Semi-Leptonic Decays\\ \>\> of Pseudoscalar mesons\\
15.00 \>  B. Moussalam \>An estimate of
the $p^4$ EM contribution to $M_{\pi^+} - M_{\pi^0}$\\
15.30 \>  S. Peris \>Matching Long and Short
Distances in large $N_c$ QCD\\
16.00 \> Coffee\\
16.40 \>  E. de Rafael \> QCD at large-$N_c$ and Weak matrix Elements\\
17.25 \>
A. Rusetsky \> Bound States with Effective Lagrangians\\
17.55 \> End of session\\[1cm]

{\bf Wednesday}\\*
9.00 \>  B. Borasoy \>Long Distance Regularization
and Chiral Perturbation\\ \>\> Theory\\
9.30 \>  J. Kambor \>Generalized Heavy Baryon
Chiral Perturbation Theory\\ \>\> and the Nucleon Sigma term\\
10.10 \>  T. Hemmert \>Baryon CHPT and
Resonance Physics\\
10.50 \> Coffee\\
11.20 \>  M. Mojzic \>Nucleon properties
to O($p^4$) in HBCHPT\\
11.50 \>  N. Fettes \> Pion-Nucleon scattering in Chiral Perturbation theory\\
12.20 \>  G. H\"ohler \>Relations of Dispersion Theory to Chiral Effective
 \\ \>\>Theories for $\pi N$ Scattering\\
13.00\> End of Session\\
14.20 \>  S. Steininger \>Isospin violation in
the Pion Nucleon System\\
14.50 \>  M. Sainio \> Goldberger-Miyazawa-Oehme
Sum Rule Revisited\\
15.20 \>  E. Epelbaoum \>
Low-Momentum Effective Theory for Two Nucleons\\
15.50 \> Coffee\\
16.20 \>  M. Savage \>Effective Field Theory
in the Two-Nucleon Sector\\
17.00 \>  N. Kaiser \>Chiral Dynamics of NN
Interaction\\
17.35 \>  S. Beane \> Compton scattering off
the Deuteron in HBCHPT\\
18.10 \> End of session\\[1cm]

{\bf Thursday}\\*
9.00 \>  H. Leutwyler \> $1/N_c$ Expansion in CHPT
and Lorentzinvariant\\ \>\> HBCHPT: Introduction\\
9.30 \>   T. Becher \> HBCHPT in a Lorentzinvariant
Form\\
10.10 \>  R. Kaiser \>Chiral $SU(3)\times SU(3)$ at large
$N_C$: The $\eta$-$\eta^\prime$ system\\
10.50 \> Coffee\\
11.30 \>  D. Diakonov \>Derivation of the Chiral
Lagrangian from the Instanton\\ \>\> Vacuum\\
12.15 \>  J. Stern \>Theoretical Study of Chiral
Symmetry Breaking:\\ \> \> Recent Developments\\
13.00\> End of Session\\
14.20 \>  G.~Colangelo \>Numerical Solutions
of Roy Equations\\
15.00 \>  J. Gasser \> The One-Channel Roy
Equation\\
15.35 \> Coffee\\
16.05 \>  G. Wanders \> How do the Uncertainties on the Input
Affect the\\ \>\> Solution of he Roy Equations ?\\
16.40 \>  J. Schechter \>Low Energy $\pi K$
Scattering and a Possible Scalar Nonet\\
17.20 \>  A. Pich \>CHPT Phenomenology in
the Large-$N_C$ Limit\\
18.00 \> End of session\\[1cm]

{\bf Friday}\\*
9.00 \>  A. Nyffeler \>Gauge Invariant Effective Field Theory for
a Heavy\\ \>\> Higgs Boson\\
9.35 \>  J. Goity \>The Goldberger-Treiman
Discrepancy in the\\ \>\>  Chiral Expansion\\
10.10\>  L. Girlanda \>Tau decays and
Generalized CHPT\\
10.35 \> Coffee Break\\
11.15 \>  F. Orellana\>Different Approaches
to Loop Calculations in\\
\>\> ChPT Exemplified with the Meson Form Factors\\
11.40 \>  D. Toublan \> From Chiral Random Matrix
Theory to Chiral\\ \>\> Perturbation Theory\\
12.10 \> J. Bijnens \> Farewell
\end{tabbing}

\section{Participants and their email}

\begin{tabbing}
A very long name\=a very long institute\=email\kill

S. Beane  \>Maryland \>sbeane@physics.umd.edu\\
T. Becher \>Bern\>     becher@itp.unibe.ch    \\
V. Bernard \>Strasbourg\> bernard@lpt1.u-strasbg.fr\\
J. Bijnens \>Lund\>   bijnens@thep.lu.se    \\
B. Borasoy \>Amherst\>borasoy@het.phast.umass.edu     \\
P. Buettiker \>FZ J\"ulich\>  p.buettiker@fz-juelich.de    \\
G. Colangelo \>Z\"urich\>  gilberto@physik.unizh.ch    \\
G. D'Ambrosio \>Naples\> dambrosio@na.infn.it  \\
D. Diakonov \>Nordita\>  diakonov@nordita.dk  \\
J. Donoghue \>Amherst\>   donoghue@phast.umass.edu  \\
G. Ecker \>Wien\> ecker@merlin.pap.univie.ac.at         \\
E. Epelbaoum \>FZ J\"ulich\> e.epelbaum@fz-juelich.de   \\
N. Fettes \>FZ J\"ulich\> n.fettes@fz-juelich.de  \\
M. Franz \>Bochum\> mario.franz@tp2.ruhr-uni-bochum.de       \\
J. Gasser \>Bern\>  gasser@itp.unibe.ch       \\
L. Girlanda \>Orsay\> girlanda@ipno.in2p3.fr     \\
K. Goeke \>Bochum\> goeke@hadron.tp2.ruhr-uni-bochum.de      \\
J. Goity \>Jefferson Lab\> goity@cebaf.gov         \\
M. Golterman \>St.Louis\>maarten@aapje.wustl.edu   \\
S. G\"usken \>Wuppertal\>guesken@theorie.physik.uni-wuppertal.de   \\
T. Hemmert \>FZ J\"ulich\>  th.hemmert@fz-juelich.de       \\
G. H\"ohler \>Karlsruhe\> gerhard.hoehler@physik.uni-karlsruhe.de  \\
N. Kaiser \>TU M\"unchen\> nkaiser@physik.tu-muenchen.de         \\
R. Kaiser \>Bern\> kaiser@itp.unibe.ch        \\
J. Kambor \>Z\"urich\> kambor@physik.unizh.ch    \\
H.-C. Kim \>Pusan\> hchkim@hyowon.pusan.ac.kr       \\
M. Knecht \>Marseille\>knecht@cpt.univ-mrs.fr    \\
H. Leutwyler \>Bern\>leutwyler@itp.unibe.ch      \\
U. Mei{\ss}ner \>FZ J\"ulich\> Ulf-G.Meissner@fz-juleich.de   \\
M. Moj\v zi\v c \>Bratislava\> Martin.Mojzis@fmph.uniba.sk\\
B. Moussallam \>Orsay\> moussall@ipno.in2p3.fr   \\
G. M\"uller \>Wien\> mueller@itkp.uni-bonn.de     \\
H. Neufeld \>Wien\>  neufeld@merlin.pap.univie.ac.at      \\
A. Nyffeler \>DESY, Zeuthen\> nyffeler@ifh.de      \\
F. Orellana \>Zurich\> fjob@physik.unizh.ch     \\
E. Pallante \>Barcelona\> pallante@greta.ecm.ub.es  \\
S. Peris \>Barcelona\> peris@ifae.es    \\
H. Petry \>Bonn\> petry@pythia.itkp.uni-bonn.de        \\
A. Pich  \>Valencia\> pich@chiral.ific.uv.es     \\
J. Prades \>Granada\> prades@ugr.es     \\
E. de Rafael \>Marseille\> EdeR@cptsu5.univ-mrs.fr    \\
A. Rusetsky \>Bern\> rusetsky@itp.unibe.ch     \\
M. Sainio \>Helsinki\>sainio@phcu.helsinki.fi     \\
M. Savage \>Seattle\> savage@phys.washington.edu     \\
J. Schechter \>Syracuse\>schechter@suhep.phy.syr.edu    \\
K. Schilling \>Wuppertal\>schillin@wpts0.physik.uni-wuppertal.de    \\
S. Steininger \>FZ J\"ulich\> s.steininger@fz-juelich.de     \\
J. Stern \>Orsay\> stern@ipno.in2p3.fr        \\
P. Talavera \>Lund\> pere@thep.lu.se \\
D. Toublan \>Stony Brook\>toublan@nuclear.physics.sunysb.edu \\
A. Vladikas \>Rome\>  vladikas@roma2.infn.it     \\
G. Wanders \>Lausanne\> gerard.wanders@ipt.unil.ch    \\
A. Wirzba \> Darmstadt\>andreas.wirzba@physik.tu-darmstadt.de
\end{tabbing}
\newabstract
\section{A short guide to review literature}

Chiral Perturbation Theory grew out of current algebra, and it  soon
was realized that certain terms beyond the lowest order were also uniquely
defined. This early work and references to earlier review papers can be found
in \cite{1}. Weinberg then proposed a systematic method in \cite{2},
later systematized
and extended to use the external field method in the classic papers
by Gasser and Leutwyler \cite{3},\cite{4}, which, according to Howard Georgi,
everybody should put under his/her pillow before he/she goes to sleep.
The field has since then extended a lot
and relatively recent review papers are: Ref.\cite{5} with an emphasis on the
anomalous sector, Ref.\cite{6} giving a general overview over the vast field
of applications in  various areas of physics, Ref.\cite{7} on mesons and
baryons,
and Ref.\cite{8} on baryons and multibaryon processes.
In
addition there are books
by Georgi\cite{9}, which, however, does not cover the standard approach,
including the terms in the lagrangian at higher order and
a more recent one by Donoghue, Golowich and Holstein\cite{10}.

There are also the lectures available on the archives by H.~Leutwyler
\cite{10b}
E.~de~Rafael \cite{11}, A.~Pich \cite{12b}, G. Ecker \cite{12c}
as well as numerous others (the single nucleon sector is covered in
most detail in \cite{12d})..
The references to the previous
meetings are \cite{13},\cite{14},\cite{15},\cite{15b}.
There are also the proceedings of the Chiral Dynamics
meetings at MIT (1994) \cite{16} and in Mainz (1997) \cite{16b}.
The DA$\Phi$NE
handbook \cite{17} also contains useful overviews. The $NN$ sector
is covered in the proceedings of the Caltech workshop\cite{18}.

\newabstract
\begin{center}
{\large\bf The Generating Functional\\[5pt] 
at Next-to-Next-to-Leading Order}\\[0.5cm]
Gerhard Ecker \\[0.3cm]
Institut f\"ur Theoretische Physik, Universit\"at Wien \\
Boltzmanng. 5, A-1090 Wien, Austria\\[0.5cm]
\end{center}

\noindent
The generating functional for Green functions of quark currents
receives contributions from tree, one-loop and two-loop diagrams 
at $O(p^6)$. After a classification of these diagrams, 
I describe the construction of the effective chiral Lagrangian of 
$O(p^6)$ for a general number $N_f$ of light flavours \cite{BCE2}. 
Special matrix relations reduce the number of independent
terms substantially for $N_f=$ 3 and 2. The $SU(3)$ Lagrangian is
compared with the Lagrangian of Fearing and Scherer \cite{FS}.

The one- and two-loop contributions to the generating functional are
divergent. In a mass independent regularization scheme like dimensional
regularization, the divergent parts are polynomials in masses and 
external momenta. The cancellation of nonlocal divergences is an
important consistency check for the renormalization procedure 
\cite{BCE2}. The double and single poles in
$d-4$ in the coefficients of the local divergences are then 
cancelled by the divergent parts of the coupling constants
of $O(p^6)$. The dependence of the renormalization procedure on the 
choice of the chiral Lagrangian of $O(p^4)$ is discussed. 

The divergence structure of the generating functional can be
used to check specific
two-loop calculations in chiral perturbation theory. Moreover,
it provides the leading infrared singular pieces of Green functions,
the  chiral double logs. These double logs
($L^2$) come together with terms of the form $L\times L_i^r$ and
products  $L_i^r\times L_j^r$, involving also the low-energy constants
$L_i^r(\mu)$ of $O(p^4)$. It is natural to include
all such terms in the analysis especially because they are often
comparable to or even bigger than the proper double-log terms.
All contributions of this type to the generating functional can be 
given in closed form \cite{BCE1}.
This generalized double-log approximation is applied to several
quantities of interest such as mesonic decay constants and form
factors where complete calculations to $O(p^6)$ are not yet available.
The results indicate where large $p^6$ corrections are
to be expected.

\newabstract

\begin{center}
{\large\bf Pion Form Factors at $p^6$}\\[0.5cm]
J. Bijnens$^1$, G. Colangelo$^2$ and {\bf P. Talavera}$^1$\\[0.3cm]
$^1$Dep. Theor. Phys. 2, Lund University,\\
S\"olvegatan 14A, S22362 Lund, Sweden\\[0.3cm]
$^2$Inst. Theor. Phys., Univ. Z\"urich,\\
Winterthurerstr. 190, CH--8057  Z\"urich--Irchel, Switzerland.\\[0.3cm]
\end{center}

We compute the 
vector and scalar form factors of the pion to
two loops in CHPT and
compare carefully with the existing data.

For the scalar form factor this involves a comparison with the form factor
derived using dispersion theory and chiral constraints from
the $\pi\pi$ phase shifts\cite{1}. The CHPT
formula fits well over the entire range of validity. Moreover, we 
show that using the ``modified Omn\`es representation'' 
which exponentiates the unitarity correction, the chiral
representation improves and follows the exact form factor up to about
700 MeV. 

For the vector form factor we collected all available data
and performed the standard simple fits.
We fit the CHPT formula at two loops
together with a phenomenological higher order term to obtain
the pion charge radius and $c_V^\pi$:
\begin{equation}
{\langle r^2\rangle^\pi_V} = (0.437\pm0.016)\mbox{ fm}^2 \; \; ,
\quad
c_V^\pi = (3.85\pm0.60)\mbox{ GeV}^{-4} \; \; .
\end{equation}
The error is a combination of theoretical and experimental errors.

By comparing to the Taylor expansions of the measured form factors,
we have been able to better determine some of 
the LEC's: $\bar{l}_4$ and $\bar{l}_6$.
$\bar{l}_6$ together with results from 
 $\pi \to l\nu\gamma$\cite{3}
lead then to $\bar{l}_5$.
\begin{equation}
\bar{l}_4 = 4.4\pm 0.3\;,\quad\bar{l}_6 = 16.0\pm0.5 \pm 0.7 \; 
\quad\mbox{and}\quad
\bar{l}_5 = 13.0 \pm 0.9 \; \; .
\end{equation}
The other two LEC's we determined are ${\cal O}(p^6)$ constants,
that contribute to the quadratic term in the polynomial of the scalar and
vector form factors. We found
\begin{equation}
r^r_{S3}(M_\rho) \simeq 1.5 \cdot 10^{-4}\; , \; \; \; \; r^r_{V2}(M_\rho)
\simeq 1.6 \cdot 10^{-4} \; \; ,
\end{equation}
with a substantial uncertainty. 
These values are rather close to those obtained with
the resonance saturation hypothesis, supporting the idea
that this hypothesis works also at order $p^6$.The full discussion
can be found in \cite{2}.

\newabstract

\begin{center}
{\large\bf Quenched and Partially Quenched Chiral Logarithms}\\[0.5cm]
{\bf Maarten Golterman}\\[0.3cm]
Department of Physics, Washington University,\\
St. Louis, MO 63130, USA\\[0.3cm]
\end{center}

In this talk, I reported on the recent quenched hadron
spectrum results of the CP-PACS collaboration
\cite{cppacs}.  I 
discussed in particular the evidence for quenched chiral logarithms
in the pion mass \cite{bergol1},\cite{sharpe1}.  In the second part, I gave an
introduction to partially quenched chiral perturbation theory
\cite{bergol2},\cite{shazha},\cite{sharpe2},\cite{golleu},
and discussed its relevance for the analysis of future results from
lattice QCD.

\newabstract

\begin{center}
{\large\bf The Generating Functional for Hadronic Weak Interactions and its 
Quenched Approximation
          }\\[0.5cm]
 {\bf Elisabetta Pallante}$^1$\\[0.3cm]
$^1$Facultat de F\'\i sica, Universitat de Barcelona,\\
Av. Diagonal 647, 08028 Barcelona, Spain.\\[0.3cm]
\end{center}

Chiral Perturbation Theory (ChPT) combined with lattice QCD is a promising
 tool for computing hadronic  weak matrix elements at long distances. 
 Here I discuss the derivation of the one loop generating functional of ChPT
  in the 
presence of weak interactions with  $\vert\Delta S\vert =1,\, 2$ 
 and the modifications induced by the {\em quenched} approximation adopted 
on the lattice.  The framework I use is known  as
 quenched ChPT \cite{qCHPT}, while its extension to hadronic weak 
interactions can be found in \cite{EP1} and refs. therein.
The advantage of deriving the generating functional is twofold:
 first, it allows
for a systematic control on the quenched modifications (i.e. how much 
the coefficients of the chiral logs are modified by quenching) and second,
 it gives in
one step the coefficients of all chiral logarithms for any Green's function or
S-matrix element, full and quenched.
%Within the full theory, the u.v. divergences of the $p^4$
%counterterms, octet and 27-plet, are derived for a generic number of flavours
% $N$.
The main relevant modification induced by quenching is the
 appearance of the so called {\em quenched chiral logs} both in the strong
 and weak sector. They  can be accounted for via a redefinition of the 
leading  order parameters associated to the mass-like terms
 (the usual mass term and the weak mass term). 
%Also, the quenched approximation largely reduces the number of u.v. divergent 
%counter-terms in the octet sector.
As an immediate application, the full and quenched behaviours of the chiral
 logarithms which appear in $K\to\pi\pi$ matrix elements can be studied both
 for $\Delta I=1/2$ and $3/2$. The numerical analysis 
shows that 
%quenching largely affects the coefficient of the chiral
%logarithm in the 27-plet $\Delta I=3/2$ and $\Delta I=1/2$ amplitudes. In
%addition, 
the modification induced by quenching follows
a pattern that tends to suppress the $\Delta I=1/2$ dominance.

\newabstract
\begin{center}
{\large\bf A Super-Heat-Kernel Representation for the One-Loop
Functional of Boson--Fermion Systems}\\[0.5cm] 
{\bf Helmut Neufeld}\\[0.3cm]
Institut f\"ur Theoretische Physik der Universit\"at Wien,\\
Boltzmanngasse 5, A-1090 Wien, Austria.\\[0.3cm]
\end{center}

The one-loop functional of a general quantum field theory with 
bosons and fermions can be written in terms of the super-determinant of a
super-matrix differential-operator. This super-determinant is then 
further evaluated by using heat-kernel methods. This approach
\cite{Berezinian} corresponds
to a simultaneous treatment of bosonic, fermionic and mixed loop-diagrams.
The determination of the one-loop divergences is reduced to simple
matrix manipulations, in complete analogy to the familiar heat-kernel
expansion technique for bosonic or fermionic loops.

Applications to the renormalization of the pion--nucleon interaction
\cite{SHK} and
to chiral perturbation theory with virtual photons and leptons
\cite{Knecht} demonstrate
the efficiency of the new method.
The cumbersome and tedious calculations of the conventional approach are
now reduced to a few simple algebraic manipulations.
The presented computational scheme is, of course, not restricted to chiral
perturbation theory, but can easily be applied or extended to any (in
general non-renormalizable) theory with boson--fermion interactions.

\newabstract

\begin{center}
{\large\bf Some Low Energy Constants from Lattice QCD: Recent Results}\\[0.5cm]
Stephan G\"usken \\[0.3cm]
Dep. Phys. , University of Wuppertal,\\
42097 Wuppertal, Germany\\[0.3cm]
\end{center}

Hadronic properties at low energies are sensitive to non-perturbative
contributions from quantum fluctuations. In particular, flavour
singlet quantities like the pion-nucleon-sigma term $\sigma_{\pi N}$
and the flavour singlet axial coupling of the proton $G_A^1$ might
be determined largely by so called disconnected insertions. These are
given by the correlation of the nucleon propagator with a
quark-antiquark vacuum loop.

Recently the SESAM collaboration has performed a full QCD lattice
simulation with $n_f=2$ dynamical fermions on 4 different values of
the sea quark mass, and with a statistics of 200 independent gauge
configurations per sea quark \cite{sesam_light_spectrum}.
In this talk we present the results of the analysis
of these gauge configurations with respect to 
$\sigma_{\pi N}$ \cite{sesam_nsigma} and 
$G_A^1$ \cite{sesam_ga}.  

SESAM finds a quite low value for the pion-nucleon-sigma term,
$\sigma_{\pi N} = 18(5)$MeV. Its smallness is directly related 
to the apparent decrease of light quark masses when unquenching lattice
QCD simulations \cite{sesam_quarkmass},\cite{cp_pacs_burkhalter}.

For the flavor singlet axial coupling of the proton, SESAM estimates
$G_A^1 = 0.20(12)$, consistent with the experimental result and with 
previous findings from quenched simulations \cite{ga1_quenched}.

\newabstract
\begin{center}
{\large\bf Quark Masses and Chiral Condensate from the Lattice}\\[0.5cm]
Anastassios Vladikas\\[0.3cm]
INFN c/o Department of Physics,
Universit\'a di Roma ``Tor Vergata'',\\
Via della Ricerca Scientifica 1,
I-00133 Rome, Italy\\[0.3cm]
\end{center}

Ward Identities can be used in order to measure, from first principles, the 
light quark masses and the chiral condensate in lattice
QCD. Particular attention is paid to the problem of chiral symmetry breaking
by the Wilson action and its restoration in the continuum limit \cite{KSBOC}.
The main sources of systematic errors in computations are: (1)
quenching; (2) finite mass extrapolation to the chiral limit; (3) scaling
violations; (4) renormalization of lattice operators in 1-loop
perturbation theory. Scaling violations can be reduced
by applying Symanzik-improvement to Wilson fermions, as
reviewed in \cite{LUESCH}. Operator renormalization
can be carried out non-perturbatively; see \cite{NP}. 

All results are in the ${\overline MS}$ scheme at renormalization scale
$\mu = 2$GeV:
\begin{eqnarray}
\langle \bar \psi \psi \rangle &=& - [245 \pm 15  MeV ]^3
\qquad \cite{APE1} \nonumber \\
\langle \bar \psi \psi \rangle &=& - [253 \pm 25 MeV ]^3
\qquad \cite{APE2} \nonumber \\
m_{u,d} &=& 5.7 \pm 0.8 MeV
\qquad \cite{APE1} \nonumber \\
m_{u,d} &=& 4.5 \pm 0.4 MeV
\qquad \cite{APE2} \nonumber \\
m_{u,d} &=& 4.6 \pm 0.2 MeV
\qquad \cite{CPPACS} \nonumber
\end{eqnarray}
Important open questions remain: (1) the dependence of the strange quark mass
on the bare mass calibration from the $\phi$- or the $K$-meson \cite{CPPACS};
(2) current unquenched quark mass results appear to be smaller by about $30\%$.

\newabstract

\begin{center}
{\large\bf Dispersive Calculation of Weak Matrix Elements}\\[0.5cm]
John F. Donoghue\\[0.3cm]
Department of Physics and Astronomy \\
University of Massachussets, Amherst, MA 01003 USA\\[0.3cm]

\end{center}
I described a project with Eugene Golowich in which we provide a 
rigorous calculational framework for certain weak non-leptonic 
matrix elements, 
valid in the chiral limit. This involves relating the weak
operators to the vacuum polarization functions of vector and
axial-vector currents. These functions obey dispersion relations and 
the inputs to these are largely known from experiment, and there are
firm theoretical constraints. Certain aspects
of this program were accomplished a few
years ago~\cite{dg1}, and we explored
the use of data for the Weinberg sum rules and for this weak
calculation~\cite{dg2}. What is new now is the understanding of how
this fits into the operator product expansion, and the separate
determination of two local operators (those related to $O_7$ and $O_8$
in the usual basis). We define these matrix elements at a scale
$\mu$ in the ${\bar{MS}}$ scheme, and verify the renormalization group
running and the OPE structure. An updated phenomenological analysis
incorporating new data was described and will be given in the upcoming
publication\cite{dg3}.

\newabstract
\begin{center}
{\large\bf $\Delta S=1$ Transitions in the $1/N_c$ Expansion }\\[0.5cm]
Johan Bijnens$^1$ and {\bf Joaquim Prades}$^2$\\[0.3cm]
$^1$Department of Theoretical Physics 2, Lund University,\\
S\"olvegatan 14A, S-22362 Lund, Sweden\\[0.3cm]
$^2$Departamento de F\'{\i}sica Te\'orica y del Cosmos, Universidad
de Granada, Campus de Fuente Nueva, E-18002 Granada, Spain\\[0.3cm]
\end{center}
In this talk we present the results obtained in a recent
work on the $\Delta I=1/2$ rule in the chiral
limit \cite{BP98}.
 In particular, we discuss the matching between long- and 
short-distance  contributions at  next-to-leading  in a $1/N_c$
expansion
and show how the scheme-dependence from the two-loop renormalization group
running can be treated. We then use this method to
study the three $O(p^2)$ couplings modulating the
terms contributing to non-leptonic 
kaon decays, namely the usual octet and 27-plet derivative terms
 as well as the weak mass term. We use
 the Extended Nambu--Jona-Lasinio model as the low energy approximation.

The known unsatisfactory high energy behaviour at large $N_c$ 
of this model
we treat as explained in \cite{BP98}. For attempts to avoid this problem
see the talks by Santi Peris and Eduardo de Rafael.
At present, their method cannot be applied to the quantities considered here.

Reasonable matching for the three $O(p^2)$ couplings introduced before
is obtained. We predict them within ranges and obtain
 a huge enhancement of the $\Delta I=1/2$ amplitude
with respect to the $\Delta I=3/2$ one. This we 
identify to come  from $Q_2$ 
and $Q_6$ Penguin-like diagrams. 
These predictions are {\em parameter free} and agree within the uncertainties
with the experimental values.

We also show how the  factorizable contributions from the $Q_6$ 
operator  are IR divergent. This divergence cancels
only when the non-factorizable contributions are added consistently. 
This makes the $B_6$ parameter not well defined.

We believe that this work presents some advances towards the 
mastering of non-leptonic kaon decays. 
This will be pursued in  forthcoming works determining
the $\Delta S=1$ non-leptonic couplings at $O(p^4)$ 
and $\varepsilon'/\varepsilon$ within the Standard Model \cite{BPP99}.

\newabstract

\begin{center}
{\large\bf Effective $\Delta S = 1$ Weak Chiral Lagrangian in 
the Instanton-induce Chiral Quark Model}\\[0.5cm]
Mario Franz$^1$, {\bf Hyun-Chul Kim}$^2$, and Klaus Goeke$^1$\\[0.3cm]
$^1$Inst. f. Theo. Phys. II, Ruhr-Universit\"at Bochum,\\
D-44780 Bochum, Germany\\[0.3cm]
$^2$Dep. of Phys., Pusan National University,\\
609-735 Pusan, The Republic of Korea\\[0.3cm]
\end{center}
In this talk, we present the recent investigation of 
the effective $\Delta S = 1$ weak chiral Lagrangian 
within the framework of the instanton-induced chiral quark 
model.  Starting from the effective four-quark operators, 
we derive the effective weak chiral action by integrating out the 
constituent quark fields.  Employing the derivative expansion, we 
are able to obtain the effective $\Delta S = 1$ weak chiral 
Lagrangian to order ${\cal O}(p^4)$.  The resulting 
${\cal O}(p^4)$ low energy constants are derived as follows~\cite{FKG}:
\begin{eqnarray}
N_{1}^{(\underline{8})} &=&
\left( -{N_c^2 M^2 \over 128 \pi^4 f_\pi^2 } + {N_c \over 8 \pi^2}
- {f_\pi^2 \over 2 M^2} \right) c_6  \;+\;
\left( -{N_c M^2 \over 128 \pi^4 f_\pi^2 } + {1 \over 8 \pi^2}
- {f_\pi^2 \over 2 N_c M^2} 
\right) c_5, \nonumber
\\
N_{2}^{(\underline{8})} &=&
{N_c \over 60 \pi^2}
\left( \left( -2 +{1 \over N_c} 3 \right) c_1 + 
\left( 3 - {1 \over N_c} 2 \right) c_2 
+ {1 \over N_c} 5 c_3 + 5 c_4 \right. \nonumber \\
&& + \; \left.
\left( - 3 + {1 \over N_c} 2 \right) c_9 + 
\left( 2 - {1 \over N_c} 3 \right) c_{10} \right),
\nonumber\\
N_{3}^{(\underline{8})} &=&
0,
\nonumber\\
N_{4}^{(\underline{8})} &=&
{N_c \over 60 \pi^2 }
\left( \left( \frac{3}{2} - {1 \over N_c} \right) c_1 
+ \left( - 1 + {1 \over N_c} \frac{3}{2} \right) c_2 
+ \frac{5}{2} c_3 + {1 \over N_c} \frac{5}{2} c_4 
\right. \nonumber \\ && \hspace{10mm} \left.
- \frac{5}{2} c_5 - {1 \over N_c} \frac{5}{2} c_6
+ \left( 1 - {1 \over N_c} \frac{3}{2} \right) c_9 
+ \left( -  \frac{3}{2} + {1 \over N_c} \right) c_{10} \right),
\nonumber\\
N_{28}^{(\underline{8})} &=&
{N_c \over 60 \pi^2}
\left(  \left( -\frac{3}{2} + {1 \over N_c} \right) c_1 
+ \left( 1 - {1 \over N_c}  \frac{3}{2} \right)  c_2 
- \frac{5}{2} c_3 -  {1 \over N_c} \frac{5}{2} c_4 
\right. \nonumber \\ && \hspace{10mm} \left.
- \frac{5}{2} c_5 - {1 \over N_c} \frac{5}{2} c_6 
+ \left( - 1 + {1 \over N_c} \frac{3}{2} \right) c_9 
+ \left(  \frac{3}{2} - {1 \over N_c} \right) c_{10} \right),
\nonumber\\ 
N_{1}^{(\underline{27})} &=&
N_{5}^{(\underline{27})} =
N_{6}^{(\underline{27})} =
N_{20}^{(\underline{27})} = 0,
\nonumber\\
N_{2}^{(\underline{27})} &=&
-N_{3}^{(\underline{27})} =
-N_{4}^{(\underline{27})} =
N_{21}^{(\underline{27})} =
{N_c \over 60 \pi^2 } \left( 1 + {1 \over N_c} \right) 
\nonumber \\
&&\times 
\left( - 3 c_1 - 3 c_2 - \frac{9}{2} c_9 - \frac{9}{2} c_{10} \right)
\nonumber.
\end{eqnarray}

\newabstract
\begin{center}
{\large \bf Rare kaon decays: $K^{+}%
\rightarrow \pi ^{+}\ell ^{+}\ell ^{-} $ \textbf{\ and } $K_{L}%
\rightarrow \mu ^{+}\mu ^{-}$}\\[0.1cm]
\textbf{G. D'Ambrosio}\\[0cm]
INFN, Sezione di Napoli
I--80134 Napoli, Italy.\\[0cm]
\end{center}
\begin{small}
\underline{$\ K^{+}\rightarrow \pi ^{+}\gamma ^{*}$} Rare kaon decays are an
important tool to test the chiral theory and establish the Standard Model
and/or its extensions. $K^{+}\rightarrow \pi ^{+}\gamma ^{*}$ starts at $%
\mathcal{O(}p^{4})$ in $\chi $PT with loops (dominated by the $\pi \pi -$cut$%
)$ and counter-term contributions \cite{r1}.\ Higher order contributions ($%
\mathcal{O(}p^{6}))$ might be large, but not completely under control since
new (and unknown) counter-term structures will appear.\ Experimentally the $%
K^{+}\rightarrow \pi ^{+}l^{+}l^{-}$ ($l$=$e,\mu )$widths are known
while the slope is known only in the electron channel. An
interesting question is the origin of the 2.2$\sigma $ discrepancy of the
ratio of the widths $e/\mu $ from the $\mathcal{O(}p^{4})$ prediction. In
Ref.~\cite{r3} we have parameterized, very generally, the $K^{+}\rightarrow 
\pi ^{+}\gamma ^{*}(q)$ form factor as 
\begin{equation}
W_{+}(z)\,=\,G_{F}M_{K}^{2}\,(a_{+}\,+\,b_{+}z)\,+\,W_{+}^{\pi \pi }(z)\;,
\end{equation}
with $z=q^{2}/M_{K}^{2}$, and where $W_{+}^{\pi \pi }(z)$ is the loop
contribution given by the unitarity cut of $K^{+}\rightarrow \pi ^{+}\pi ^{+}%
\pi ^{-}$. The two unknown parameters $a_{+}$ and $b_{+}$ can be fixed from
rate and the slope in the electron channel to predict the muon rate and
consequently the ratio of the widths $e/\mu ,$ which still comes out 2.2$%
\sigma $ away from the expt. result, pointing out either an experimental
problem in the $\ $muon channel or a violation of lepton universality. Also
we speculate on the prediction of Vector Meson Dominance (VMD) for this
channel.

\underline{\textit{$K_{L}\rightarrow \mu ^{+}\mu ^{-}$.}} To fully exploit
the potential of $K_{L}\rightarrow \mu ^{+}\mu ^{-}$ in probing
short--distance dynamics it is necessary to have a reliable control on its
long--distance amplitude. The branching ratio can be generally decomposed as 
$B(K_{L}\rightarrow \mu ^{+}\mu ^{-})\,=\,|{\Re e\mathcal{A}}|^{2}\,+\,|{\Im
m\mathcal{A}}|^{2}$ and ${\Re e\mathcal{A}}\,=\,{\Re e\mathcal{A}}%
_{long}\,+\,{\Re e\mathcal{A}}_{short}$. The recent measurement of $%
B(K_{L}\rightarrow \mu ^{+}\mu ^{-})$  is almost saturated by the absorptive
amplitude leaving a very small room for the dispersive contribution~: $|{\Re
e\mathcal{A}}_{exp}|^{2}\,=\,(-1.0\pm 3.7)\times 10^{-10}$. 
Within the Standard Model the NLO short-distance amplitude  gives the
possibility to extract a lower bound on $\overline{\rho },$ once we have
under control the dispersive contribution generated by the two--photon
intermediate state. In order to saturate this lower bound we propose \cite
{DIP} a low energy parameterization of the $K_{L}\rightarrow \gamma
^{*}\gamma ^{*}$ form factor that include the poles of the lowest vector
meson resonances. Using experimental slope from $K_{L}\rightarrow \gamma
\ell ^{+}\ell ^{-}$ and  QCD constraint we predict $\overline{\rho }%
>-0.38\;(90\%C.L.)$. This  bound could be very much improved if the linear
and quadratic slope of the $K_{L}\rightarrow \gamma ^{*}\gamma ^{*}$ form
factor  were measured with good precision and a more stringent bound on $|{%
\Re e\mathcal{A}}_{exp}|$ is established.
\\[-1cm]

\end{small}

\newabstract

\begin{center}
{\large\bf Electromagnetic Corrections to Semi-leptonic}\\
{\large\bf Decays of Pseudoscalar Mesons}\\[0.5cm]
{\bf{M. Knecht}}$^1$,
H. Neufeld$^2$, H. Rupertsberger$^2$ and P. Talavera$^3$\\[0.3cm]
$^1$Centre de Physique Th\'eorique,\\ 
CNRS-Luminy, Case 907, F-13288 Marseille Cedex 9, France\\[0.3cm]
$^2$Inst. f. Theor. Phys. der Universit\"at Wien,\\
Boltzmanngasse 5, A-1090 Wien, Austria.\\[0.3cm]
$^3$Dept. Theor. Phys., Lund University,\\
S\"olvegatan 14A, S-22362 Lund, Sweden.
\end{center}

In order to extract the information on hadronic matrix elements of 
QCD currents made of light quarks from 
high-statistics semi-leptonic decays of pseudoscalar mesons, a 
quantitative understanding of electromagnetic corrections to these 
processes is necessary. Virtual photons spoil the factorization 
property of the effective Fermi theory, so that the description 
of radiative corrections must be done within an extended framework, 
which includes also the leptons. An effective theory for the interactions 
of the light pseudoscalar mesons with light leptons and with photons, 
and which 
respects all the properties required by chiral symmetry, has been 
constructed. It is based on a power counting consistent with the loop 
expansion. The renormalization of this effective theory has been studied at 
the one-loop order. The divergences arising from the loops involving a lepton 
require, besides the usual mass and wave function renormalizations, 
only three additional nontrivial counter-terms. Applications to the 
$\pi_{\ell 2}$ and pion beta decays are being completed. Further work will 
also consider the semi-leptonic decays of the kaons, in view of the forthcoming high-precision data  from the KLOE experiment at the DAPHNE $\phi$--Factory. The case of the $K_{\ell 4}$
decays of charged kaons with two charged pions in the final state
is of particular importance, since they allow 
to access the $\pi$--$\pi$ phase-shifts at low energies.

\newabstract

\begin{center}
{\large\bf An estimate of the O(p$^4$) E.M. contribution to
M$_{\pi^+}$-M$_{\pi^0}$  }\\[0.5cm]
Bachir Moussallam\\[0.3cm]
Groupe de Physique Th\'eorique, IPN,\\
Universit\'e Paris-Sud, 91406 Orsay\\[0.3cm]
\end{center}

We reanalyze a sum rule due to Das et al.\cite{dgmly}. This sum rule is
interesting not as an approximation to the physical $\pi^+-\pi^0$ mass
difference but as an exact result for a chiral low-energy parameter. A
sufficiently precise evaluation provides a model independent estimate for a
combination of $O(p^4)$ electromagnetic chiral low-energy parameters recently
introduced by Urech\cite{urech}. Three ingredients are necessary in order to
reach the required level of accuracy: firstly one must use a euclidian space
approach, secondly one must use accurate experimental data such as provided
recently by the ALEPH collaboration on $\tau$ into hadrons
decays\cite{aleph}. Finally, it is necessary to extrapolate to the chiral
limit $m_u=m_d=0$. We show how a set of sum rules allows to perform this 
extrapolation in a reliable way\cite{preprint}.

\newabstract

\begin{center}
{\large\bf Matching Long and Short Distances
             \\[0.5cm] in Large-$N_c$ QCD}\\[0.5cm]
{\bf Santiago Peris}$^1$ , Michel Perrottet$^2$ and 
Eduardo de Rafael$^2$\\[0.3cm]
$^1$Grup de Fisica Teorica and IFAE,\\
Universitat Autonoma de Barcelona, 08193 Barcelona, Spain\\[0.3cm]
$^2$Centre de Physique Theorique, CNRS-Luminy, Case 907\\
F-13288 Marseille Cedex 9, France\\[0.3cm]
\end{center}

It is shown, with the example of the experimentally known  Adler function, 
that there is no matching in the intermediate region between the two
asymptotic regimes described by perturbative QCD (for the very
short--distances) and by chiral perturbation theory (for the very
long--distances).  We then propose to consider an approximation to 
large--$N_c$ QCD which consists in restricting the hadronic spectrum in the
channels with $J^P$ quantum numbers
$0^{-}$, $1^{-}$, $0^{+}$ and $1^{+}$  to the lightest
state and in treating the rest of the narrow states as a
perturbative QCD continuum;  the onset of this continuum being fixed by
consistency constraints from the operator product expansion. We show how to
construct the low--energy effective Lagrangian which describes this
approximation. A comparison of the corresponding
predictions, to ${\cal O}(p^4)$ in the chiral expansion, with the
phenomenologically known couplings $L_i$ is also made in terms of a 
{\it single} unknown, namely $f_{\pi}/M_V$ \cite{LMD}:
\begin{equation}
\rlabel{els}
6 L_1 = 3 L_2 = \frac{-8}{7} L_3 = 4 L_5 = 8 L_8 = \frac{3}{4} L_9 = - L_{10}
= \frac{3}{8} \frac{f_{\pi}^2}{M_V^2}\, ,
\end{equation}  
where $f_{\pi}, M_V$ are the pion decay constant and the $1^-$ state's mass,
respectively.

\newabstract

\begin{center}
{\large\bf Large $N_c$ QCD and Weak Matrix Elements}\\[0.5cm]
{\bf Eduardo de Rafael}\\[0.3cm]
 Centre  de Physique Th{\'e}orique\\
       CNRS-Luminy, Case 907\\
    F-13288 Marseille Cedex 9, France\\[0.3cm]
\end{center}

The first part of my talk was an overview of the progress made and the
problems which remain in deriving an effective
Lagrangian which describes the non--leptonic weak interactions of the
Nambu--Goldstone degrees of freedom ($K$,
$\pi$ and $\eta$) of the  spontaneous $SU(3)_{L}\times SU(3)_{R}$ symmetry
breaking in the Standard Model. I showed, with examples, how the
coupling constants of the $\vert\Delta S\vert=1$ and $\vert\Delta
S\vert=2$  effective low energy Lagrangian are given by {\it integrals} of
Green's functions of QCD currents and density currents, while those of the
strong sector (QCD only) are more simply related to the coefficients of
the  Taylor expansions of QCD Green's functions. The study of these issues
within the framework of the $1/N_c$ expansion, and in the {\it lowest meson
dominance} approximation described in ref.~\cite{PPdeR98}, seems a very
promising path.

The second part of my talk was dedicated to a review of the properties of
the
$\Pi_{LR}(Q^2)$ correlation function in the large--$N_c$
limit as recently discussed in ref.~\cite{KdeR97}.
This is the correlation function which governs the {\it electroweak}
$\pi^{+}-\pi^{0}$ mass difference. Following the discussion of
ref.~\cite{KPdeR98}, I showed how the calculation of this observable,
which requires non--trivial contributions from next--to--leading terms
in the $1/N_c$ expansion, provides an excellent theoretical laboratory
for studying issues of long-- and short-- distance matching in
calculations of weak matrix elements of four--quark operators.

The third part of my talk was dedicated to recent work reported in
ref~\cite{KPdeR98d}, where
it is shown that the $K\rightarrow\pi\pi$ matrix elements of the four--quark
operator
$Q_7$, generated by the electroweak penguin--like diagrams of the Standard
Model, can be calculated to first non--trivial order in the chiral
expansion and in the
$1/N_c$ expansion. I compared the results to recent numerical
evaluations from lattice--QCD.

\newabstract

\begin{center}
{\large\bf
Bound States with Effective Lagrangians: \\
Energy Level Shift in the External Field}\\[0.5cm]
Vito Antonelli$^1$, Alex Gall$^1$, J\"{u}rg Gasser$^1$ and 
{\bf Akaki Rusetsky}$^{1,2,3}$\\[0.2cm]
$^1$Institute for Theoretical Physics, University of Bern,\\
Sidlerstrasse 5, CH-3012, Bern, Switzerland\\[0.2cm]
$^2$Laboratory of Theoretical Physics, JINR, 141980 Dubna, Russia\\[0.2cm]
$^3$IHEP, Tbilisi State University, 380086 Tbilisi, Georgia\\[0.2cm]
\end{center}

Recent growth of interest in both the
experimental and theoretical study of the properties of hadronic atoms
is motivated by the possibility of direct 
determination of the strong hadronic scattering lengths from the atomic
data.
The detailed analysis of the $\pi^+\pi^-$ atom decay 
within ChPT has been carried out under the assumption of locality of strong 
interactions at the atomic length scale~\cite{Atom}.
Two conceptual difficulties arise in the theory beyond this approximation:

\noindent $\bullet$ 
Relativistic approaches to the bound-state problem
deal with the off-shell Green's functions. It is at present unclear whether
the ambiguity of the off-shell extrapolation in the effective theory
affects the bound-state observables. 

\noindent $\bullet$ 
The bound-state observables in nonrenormalizable theories
contain additional UV divergences in the matrix elements of strong 
amplitudes between the bound-state wave functions.
It is at present unclear whether these divergences can be
canceled by the same LEC's which render finite the amplitudes itself.

Addressing these problems, a simple model of a heavy massive scalar
particle~- bound in an external Coulomb field~- is considered within the
nonrelativistic effective Lagrangian approach~\cite{Nonrel}. 
Radiative  corrections to the bound state energy
levels due to the interaction with a dynamical scalar "photon", are
calculated. In the model studies it is demonstrated that~\cite{New}:

\noindent $\bullet$
The ambiguity in the off-shell extrapolation of the Green's function in the
relativistic theory does not affect the bound-state spectrum.

\noindent $\bullet$
Bound-state observables are made finite by the same counter-terms which render
finite Green's functions, even in effective nonrenormalizable theories.

\noindent $\bullet$
UV divergences in the nonrelativistic bound-state matrix elements
are correlated by matching and cancel.

\newabstract

\begin{center}
{\large\bf SU(3) Baryon Chiral Perturbation Theory and 
Long Distance Regularization}\\[0.5cm]
Bu{\=g}ra Borasoy\\[0.3cm]
Department of Physics and Astronomy\\
University of Massachusetts\\
Amherst, MA 01003, USA\\[0.3cm]
\end{center}

Baryon chiral perturbation theory as conventionally applied has a well-known
problem with the SU(3) chiral expansion: loop diagrams generate very
large SU(3) breaking corrections and greatly upset the subsequent 
phenomenology. This problem is due to the portions of loop integrals that 
correspond to propagation at short distances for which the effective
theory is not valid. One can reformulate the theory just as rigorously
by regulating the loop integrals using a momentum-space cutoff which removes
the spurious short distance physics \cite{dh},\cite{dhb}.
The chiral calculations can now provide a model
independent realistic description of the very long distance physics.
In \cite{b1} and \cite{b2} this scheme is applied to the
sigma-terms and baryon axial currents. The results are promising and show
that this development may finally allow realistic phenomenology to be
accomplished in SU(3) baryon chiral perturbation theory.

\newabstract
\begin{center}
{\large\bf Generalized Heavy Baryon Chiral Perturbation Theory and 
the Nucleon Sigma Term}\\[0.5cm]
Robert Baur and {\bf Joachim Kambor}\\[0.3cm]
Institut f\"ur Theoretische Physik, Universit\"at Z\"urich,\\
Winterthurerstr. 190, 8057 Z\"urich, Switzerland\\[0.3cm]
\end{center}

The scenario of spontaneous chiral symmetry breakdown with small or
vanishing quark condensate \cite{FSS90} and it's phenomenological consequences
for the $\pi N$--system are investigated. \cite{BK98}
Standard Heavy Baryon Chiral Perturbation Theory is extended 
in order to account for the modified chiral counting rules of quark 
mass insertions. The effective lagrangian is given to
${\cal O}(p^2)$ in its most general form and to ${\cal O}(p^3)$ in the
scalar sector. As a first application, mass- and wave-function
renormalization as well as the scalar form factor of the nucleon is
calculated to ${\cal O}(p^3)$. The result is compared to a dispersive
analysis of the nucleon scalar form factor \cite{GLS91}, adopted to the 
case of a small
quark condensate. In this latter analysis, the shift of the scalar
form factor between the Cheng-Dashen point and zero momentum transfer 
is found to be enhanced
over the result assuming strong quark condensation by up to a
factor of two, with substantial deviations starting to be visible for
$r=m_s/\hat{m}\leq 12$.\cite{BK98} As a result, the nucleon sigma term as determined 
from $\pi N$-scattering data decreases
with decreasing quark condensate. \cite{Kam99} On the other hand, the 
sigma term can also be determined from the baryon masses. To leading 
order in the quark masses, $\sigma_N$ is in proportion to 
$(r-1)^{-1}(1-y)^{-1}$,
{\it i.e.} strongly increasing with decreasing quark condensate. 
If the the strange quark content of the nucleon, $y$, were known, strong 
bounds on the light quark condensate would follow. 

We also consider the so called backward sum rule for the 
difference of electric and magnetic polarizabilities of the nucleon. 
Although the effect of a small quark condensate is less pronounced here,
this observable has the advantage of being directly experimentally 
accessible. Detailed numerical study of this sum rule is under way.

\newabstract

\begin{center}
{\large\bf Heavy Baryon ChPT and Nucleon Resonance Physics}\\[0.5cm]
Thomas R. Hemmert \\[0.3cm]
FZ J{\" u}lich, IKP (Th), D-52425 J{\" u}lich, Germany\\[0.3cm]
\end{center}

Several calculations \cite{delta}
have appeared since the ``small scale expansion'' (SSE) has been
presented to the chiral community at the Trento workshop in 1996
\cite{trento},\cite{letter}.
The idea is to incorporate the effects of low lying nucleon resonances via
a phenomenologically motivated power counting in ${\cal O}(\epsilon^n)$ with
$\epsilon=\{m_\pi,p,\Delta\}$, which supersedes the standard ${\cal O}(p^n)$
power counting of HBChPT. The new scale $\Delta$ corresponds to the energy
difference between the mass of a resonance and the mass of the nucleon and in
SSE is
counted as a small parameter\footnote{In ChPT the nucleon-delta mass splitting
counts as ${\cal O}(p^0)={\cal O}(1)$, in contrast to SSE.} of ${\cal
O}(\epsilon^1)$.
SSE therefore not only allows for calculations with explicit resonance degrees
of
freedom but also resums the resonance effects into lower orders of the
perturbative
expansion---for example see the discussion regarding the spin-polarizabilities
of the
nucleon in \cite{menu}. The complete ${\cal O}(\epsilon^2)$ SSE lagrangians for
$NN$,
$N\Delta$ and $\Delta\Delta$ are now worked out and published \cite{HHK}.
Progress
has also been achieved at ${\cal O}(\epsilon^3)$ where the complete lagrangian
for
single nucleon transitions has been worked out \cite{bfhm}. The
${\cal O}(\epsilon^3)$ divergence structure was found to be quite
different from the corresponding ${\cal O}(p^3)$ one in HBChPT. In addition to
modifications in the beta-functions of the 22 HBChPT counter terms (c.t.s)
\cite{Ecker} one needs 10 extra c.t.s to account for additional divergences
proportional to the new scale $\Delta^n,\, n\leq 3$. The finite parts of theses
10 c.t.s are utilized to guarantee a smooth transition from ${\cal
O}(\epsilon^n)$
SSE to ${\cal O}(p^n)$ HBChPT for any process in the decoupling limit
$m_\pi /\Delta\rightarrow 0$.

\newabstract

\begin{center}
{\large {\bf Nucleon properties to }}${\bf O(p}^4{\bf )}$ {\large {\bf in
HBCHPT}}\\[0.5cm]Martin Moj\v zi\v s$^1$\\[0.3cm] $^1$Dep. Theor. Phys.,
Comenius University,\\ Mlynsk\'a dolina F2, SK-84215 Bratislava, Slovakia%
\\[0.3cm]
\end{center}

Complete one-loop calculations of nucleon properties in CHPT require

\noindent 1. construction of the effective Lagrangian up to the 4th order

\noindent 2. renormalization of $m_N$, $Z_N$ and $g_A$ (and higher order
LECs)

\noindent 3. calculation of nucleon form-factors, cross-sections, $\ldots $

\smallskip\ 

\noindent A brief summary of progress made in these three points:

\smallskip\ 

\noindent 1. A {\em Mathematica} program for the construction of the
effective Lagrangian was developed. It reproduces successfully the known
results in the 2nd and 3rd orders, but does not eliminate all the dependent
terms yet (e.g. in the 3rd order it gives two terms more than it should).
The number of the 4th order terms produced by the program is $155$ so far,
this number will be probably slightly decreased in the final version.

\smallskip\ 

\noindent 2. Nucleon wave-function renormalization is known to be a tricky
issue already at the 3rd order of HBCHPT. At the 4th order yet some new
subtleties enter, but all of them have become well understood recently. The
discussion of general aspects of this topic, as well as explicit calculation
of $m_N$, $Z_N$ and $g_A$, is to be found in \cite{KM99} and references
therein.

\smallskip\ 

\noindent 3. Work in progress.

\newabstract

\begin{center}
{\large\bf Pion--nucleon scattering in heavy baryon chiral perturbation
theory}\\[0.5cm]
{\bf Nadia Fettes}$^1$, Ulf-G. Mei\ss ner$^1$ and Sven Steininger$^2$\\[0.3cm]
$^1$Forschungszentrum J\"ulich, Institut f\"ur Kernphysik (Theorie)\\
D-52425 J\"ulich, Germany\\[0.3cm]
$^2$Universit\"at Bonn, Institut f\"ur Theoretische Kernphysik\\
Nussallee 14-16, D-53115 Bonn, Germany\\[0.3cm]

\end{center}

We discuss in detail pion--nucleon scattering in the framework
of heavy baryon chiral perturbation theory to third order in small momenta.
In particular we show that the $1/m$ expansion of
the Born graphs calculated relativistically can be recovered exactly in
the heavy baryon approach without any additional
momentum-dependent wave function renormalization.
Since the normalization factor of the nucleon spinors, appearing in the
relativistic calculation, enters the heavy baryon amplitude via the wave
function renormalization\cite{smf}, we do not expand this factor.

The pion--nucleon scattering amplitude is given in terms of
four second order LECs and five combinations of LECs from ${\cal L}_{\pi
N}^{(3)}$.
In order to fix these constants,
we fit various empirical phase
shifts for pion laboratory momenta between 50 and 100 MeV.
As input we use the phase shifts of the
Karlsruhe group\cite{ka85} (KA85), from the analysis of
Matsinos\cite{em98} (EM98) and the latest update of the VPI group\cite{sp98}
(SP98).
This leads to a satisfactory description of
the phase shifts up to momenta of about 200 MeV.
The two S-waves are reproduced very well, whereas in the P-waves
the tail of the Delta is strongly underestimated and the
bending from the Roper resonance cannot be accounted for.
We also  predict threshold parameters, which turn out to be in good
agreement with analyses based on dispersion relations.
Finally we consider sub-threshold parameters
and give a short comparison to other calculations\cite{bkm},\cite{mm}.

\newabstract

\begin{center}
{\large\bf Relations of Dispersion Theory \\
to Chiral Effective Theories for $\pi N$ Scattering}\\[0.5cm]
Gerhard H\"ohler \\[0.3cm]
Institut f\"ur Theoretische Teilchenphysik der Universit\"at\\
D-76128 Karlsruhe, Germany\\[0.3cm]
\end{center}

\bigskip
Predictions for $\pi N$ scattering amplitudes within the Mandelstam triangle
and near its boundaries were recently derived from chiral perturbation 
theory (CHPT)\cite{Bernard},\cite{Mojzis}. The results were compared with
predictions from partial wave analyses or from analytic continuations
using various dispersion relations. 

Mandelstam analyticity is assumed for two steps:

\noindent
i) for constraints which lead to a {\bf unique} result of partial wave 
analysis. E. Pietarinen's expansion method (references in Ref.\cite{LB}) made it
possible to include all data available at that time in the Karlsruhe-Helsinki
solution KH80.-- The solution SP98 of R.A. Arndt et al.\cite{Arndt} is based 
on an {\em empirical parametrization} which ignores well known nearby l.h. cuts.
It includes new data, but covers only the range up to 2.1 GeV. The attempt
to apply approximately the much weaker constraint from fixed-t analyticity was 
successful for all invariant amplitudes only up to 0.6 GeV, so above this energy
the solution is questionable. 

\noindent
ii) For the continuation into the unphysical region we have used many single
variable dispersion relations, which follow from Mandelstam 
analyticity\cite{LB}. The compatibility of these relations with KH80 supports
the assumptions on the analytic properties.

\noindent
The comparison of the predictions from CHPT and dispersion relations should
be made not only for the numerical results but {\em mainly for the theoretical
expressions}. Plots of the integrands of fixed-t dispersion relations give
an important information on sub-threshold coefficients. A paper with many details
can be obtained from the author \footnote{E-mail 
gerhard.hoehler@physik.uni-karlsruhe.de}.

\newabstract
\begin{center}
{\large\bf Isospin Violation in Pion--Nucleon Scattering }\\[0.5cm]
Nadia Fettes$^1$, Ulf-G. Mei{\ss}ner$^1$ 
and {\bf Sven Steininger}$^1$,$^2$\\[0.3cm]
$^1$IKP(Theorie), Forschungszentrum J\"ulich,\\
D-52425 J\"ulich, Germany.\\[0.3cm]
$^2$ITKP, Uni Bonn,\\
Nussallee 14-16, D-53115 Bonn, Germany.\\[0.3cm]
\end{center}

We construct the complete effective chiral pion-nucleon Lagrangian in the
presence of virtual photons to one loop. 
Taking only the charged to neutral pion and the proton to neutron mass 
differences into account, 
we calculate the scattering lengths of all physical
$\pi N$--scattering processes. Furthermore we construct
six independent relations among the physical scattering amplitudes, which
vanish in the case of perfect isospin symmetry. If we now take these
relations at threshold, we find large violation in the isoscalar ones:

\begin{eqnarray}
R_1 & = & 
2 \, \frac{T_{\pi^+ p \to \pi^+ p} + T_{\pi^- p \to \pi^- p} + 2 \, T_{\pi^
0 p \to \pi^0 p}} 
          {T_{\pi^+ p \to \pi^+ p} + T_{\pi^- p \to \pi^- p} - 2 \, T_{\pi^
0 p \to \pi^0 p}} = 36 \%
\nonumber \\[0.3em] 
R_6 & = & 
2 \, \frac{T_{\pi^0 p \to \pi^0 p} - T_{\pi^0 n \to \pi^0 n}} 
          {T_{\pi^0 p \to \pi^0 p} + T_{\pi^0 n \to \pi^0 n}} = 19 \% 
\nonumber 
\end{eqnarray}

In 1977 Weinberg has already pointed out the possible large value for $R_6$
but there isn't (and probably will never be) any experimental data to verify 
this
relation. On the other hand there is hope to measure the $\pi^0 p$ scattering 
length so that one could check the prediction of the huge violation in the 
comparison
of elastic scattering of charged with neutral pions. The not mentioned 
remaining four
relations involving the isoscalar amplitudes are small (less than 2\%).

\newabstract

\begin{center}
{\large\bf Goldberger-Miyazawa-Oehme Sum Rule Revisited}\\[0.5cm]
M.E. Sainio\\[0.3cm]
Dept. of Physics, University of Helsinki,\\
P.O. Box 9, FIN-00014 Helsinki, Finland
\end{center}

There has been recently quite a lot of activity attempting to determine
the value of the pion-nucleon coupling constant with high precision.
The results vary roughly in the range $f^2=0.075-0.081$ which covers the
previous
standard value, $f^2=0.079$, but the tendency is to move the central value
downwards.

The Goldberger-Miyazawa-Oehme sum rule \cite{GMO}, the forward dispersion
relation for the $C^-$ 
amplitude at the physical threshold, provides a relationship between the
isovector $s$-wave 
pion-nucleon scattering length, the total cross section data and the pion
pole term. 
The coupling constant can be extracted ($x=\mu/m$, where $\mu$ and $m$ are
the masses for the 
charged pion and the proton)
\begin{displaymath}
f^2=\frac{1}{2}[1-(\frac{x}{2})^2]\times [(1+x)\mu a^-_{0+}-\mu^2 J^-] 
\; \; {\rm with} \; \;
J^-=\frac{1}{2\pi^2}\int_0^{\infty}\frac{\sigma^-(k)}{\omega}dk,
\end{displaymath}
where $\sigma^-=\frac{1}{2}(\sigma_- - \sigma_+)$ in 
terms of the $\pi^-p$ and $\pi^+p$ total cross sections. 

The isovector scattering length is accessible in pionic hydrogen experiments.
Also, there are plenty of cross section data up to about 350 GeV/c, which 
allows
for a determination of the integral $J^-$. However, the data around the 
$\Delta$-resonance
will have a significant influence on the $f^2$ value \cite{LS}. The precision 
of the input
information is not, at present, high enough to be able to compete with other 
methods
of determining the pion-nucleon coupling constant, but in principle the GMO 
approach can relate
more directly the uncertainty in $f^2$ to the uncertainties in experimental
quantities.

\newabstract

\begin{center}
{\large\bf Low--momentum effective theory for two nucleons}\\[0.5cm]
{\bf E.~Epelbaoum}$^{1, 2}$, W.~Gl\"ockle$^1$
and  Ulf-G. Mei\ss ner$^2$\\[0.3cm]
$^1$Ruhr--Universit\"at Bochum, Institut f\"ur theoretische Physik II\\
D--44870 Bochum, Germany\\[0.3cm]
$^2$Forschungszentrum J\"ulich, Institut f\"ur Kernphysik (Theorie)\\
D--52425 J\"ulich, Germany\\[0.3cm]
\end{center}

For the case of a Malfliet--Tjon type model potential 
\cite{1}
we show explicitly that it is possible to construct
a low--momentum effective theory for two nucleons.
To that aim we decouple the low and high momentum
components of this two--nucleon potential
using the method of unitary transformation
\cite{2}.
We find the corresponding unitary operator for the $s$--wave numerically 
\cite{3}.
The $S$--matrices in the full space and in the subspace 
of low momenta are shown to be identical.
This is also demonstrated numerically by solving the corresponding
Schr\"odinger equation in the small momentum space.

Using our exact effective theory we address some issues related
to the effective field theory approach of the two--nucleon
system
\cite{4}. 
In particular, we consider the $n p$ $^3 S_1$ and $^1 S_0$ channels.
Expanding  the heavy repulsive meson
exchange of the effective potential in a series of local contact terms
we discuss the question of naturalness of the corresponding coupling
constants. We demonstrate 
that the quantum averages of the local expansion of the effective potential
converge.
This indicates that our effective theory
has a systematic power counting. However terms of
rather high order should be kept in the effective potential to obtain
an accurate value of the binding energy (scattering length) in the $^3 S_1$ 
($^1 S_0$) channel.

We hope that this study might be useful for the real case
of NN--forces derived from chiral Lagrangians in the 
low--momentum regime.

\newabstract

\def\si{{}^1\kern-.14em S_0}
\def\siii{{}^3\kern-.14em S_1}
\def\Journal#1#2#3#4{{#1} {\bf #2}, #3 (#4)}
\def\NPB{{\em Nucl. Phys.} B}
\def\NPA{{\em Nucl. Phys.} A}
\def\NP{{\em Nucl. Phys.} }
\def\PLB{{\em Phys. Lett.} B}
\def\PRL{\em Phys. Rev. Lett.}
\def\PRD{{\em Phys. Rev.} D}
\def\PRC{{\em Phys. Rev.} C}
\def\PRA{{\em Phys. Rev.} A}
\def\PR{{\em Phys. Rev.} }
\begin{center}
{\large\bf Effective Field Theory in the Two-Nucleon Sector}\\[0.5cm]
Martin  J. Savage \\[0.3cm]
Dept. of Physics, University of Washington,  
Seattle, WA 98915, USA.\\[0.3cm]

\end{center}

The two-nucleon sector contains length scales that are much 
larger than one would naively expect from QCD.
The s-wave scattering lengths
$a^{(\si)} = -23.7\ {\rm fm}$ and $a^{(\siii)} = 5.4\ {\rm fm}$
are much greater than both
$1/\Lambda_\chi\sim 0.2\ {\rm  fm}$
and $1/f_\pi\sim 1.5\ {\rm  fm}$.
The lagrange density describing such interactions
consists of local four-nucleon operators and nucleon-pion
interactions.
Weinberg\cite{Weinberg1} suggested expanding the
NN potential in powers of the quark mass 
matrix  and external momenta.
Several phenomenological applications of this counting have 
been explored (e.g. \cite{Bira}).
However, one can show that leading
order graphs require counter-terms that occur at higher orders 
in the expansion.
A power counting was suggested\cite{KSW} that does not suffer from
these problems.
The leading interaction is the four-nucleon interaction, 
with pion exchange being sub-leading order.
Recently, Mehen and Stewart\cite{MehStew} have suggested that the 
scale for the breakdown
of the theory is $\sim 500\ {\rm MeV}$.
The deuteron can be simply incorporated into the theory and 
its moments, form factors\cite{KSW2} and polarizabilities\cite{CGSSpol}
have been computed.
The cross section for $\gamma$-deuteron Compton
scattering\cite{CGSScompt}
has been computed in this theory and agrees well with the available
data.

\newabstract

\begin{center}
{\large\bf Peripheral NN-Scattering and Chiral Symmetry}\\[0.5cm]
Norbert Kaiser\\[0.3cm]
Physik Department T39, TU M\"unchen, D-85747 Garching, Germany.\\[0.3cm]
\end{center}
\medskip

We evaluate in one-loop chiral perturbation theory all $1\pi$- and 
$2\pi$-exchange contributions to the elastic NN-interaction. We find that the 
diagrams with virtual $\Delta(1232)$-excitation produce the correct amount of 
isoscalar central attraction as needed in the peripheral partial waves ($L\geq
3$). Thus there is no need to introduce the fictitious scalar isoscalar 
$\sigma$-meson. We also compute the two-loop diagrams involving
$\pi\pi$-interaction (so-called correlated $2\pi$-exchange). Contrary to 
common believe these graphs lead to a negligibly small and repulsive 
NN-potential. Vector meson ($\rho$ and $\omega$) exchange becomes important 
for the F-waves above $T_{\rm lab}=150$~MeV. Without adjustable parameters we 
reproduce the empirical NN-phase shifts with $L\geq 3$ and mixing angles with 
$J\geq 3$ up to $T_{\rm lab}=350$~MeV. Further details on the subject can be 
found in refs.\cite{kbw},\cite{kgw}.

\newabstract

\begin{center}
{\large\bf Compton Scattering on the Deuteron in HBChPT}\\[0.5cm]
Silas R. Beane\\[0.3cm]
Department of Physics,\\
University of Maryland,\\
College Park, MD 20742 USA\\[0.3cm]
\end{center}

There exists a systematic procedure for computing nuclear processes
involving an external pionic or electromagnetic probe at energies of
order $M_\pi$\cite{weinberg}. A perturbative kernel is calculated in
baryon HBChPT and folded between phenomenological nuclear
wavefunctions.  A computation of $\pi^0$ photoproduction on the
deuteron has been performed to $O(q^4)$ in HBChPT\cite{photo}. This
process is of interest because an accurate measurement of the deuteron
electric dipole amplitude (EDA) allows a model independent extraction
of the neutron EDA, to this order in HBChPT.  Recent experimental
results from SAL indicate a value for the deuteron EDA which is very
close to the HBChPT prediction, which in turn implies a large neutron
EDA.  In similar spirit, experimental information about the
$\pi$-nucleus scattering lengths can be used to learn about $\pi$-$N$
scattering lengths \cite{weinberg}\cite{bblm}.  A recent accurate
measurement of the $\pi$-deuteron scattering length constrains HBChPT
counter-terms which contribute to the isoscalar S-wave $\pi$-$N$
scattering length, which is a particularly problematic observable.

Compton scattering on the deuteron has been computed to order $O(q^3)$
in HBChPT\cite{compton}. Our predictions at low energies are in
agreement with old Illinois data.  This process has been measured at
SAL at higher photon energies ($95MeV$) and data is currently being
analyzed. An ingredient of the deuteron process is the neutron
polarizability, which cannot be directly measured. Thus our
calculation provides a systematic means of learning about the neutron
polarizability.

\newabstract

\begin{center}
{\large\bf (a) Baryon $\chi$PT and (b) Mesons at Large $N_c$}\\[0.5cm]
H. Leutwyler\\[0.3cm]
Institute for Theoretical Physics, University of Bern,\\
Sidlerstr. 5, CH-3012 Bern, Switzerland\\[0.3cm]
\end{center}
I gave a brief introduction to some of the work being 
carried out at Bern:\footnote{see
also the reports of Thomas Becher and Roland Kaiser 
in these mini-proceedings.}\\
(a) Thomas Becher and I have shown that the approach of
Tang and Ellis \cite{Tang Ellis} can be developed into a coherent, 
manifestly Lorentz
invariant formulation of B$\chi$PT that preserves chiral power counting.
The method avoids some of the shortcomings of HB$\chi$PT. In particular, 
it allows us to analyze the low energy structure also in cases where the
straightforward nonrelativistic expansion of the amplitudes in 
powers of momenta and
quark masses fails. As an illustration, I discussed the application of the
method to the $\sigma$-term and to the 
corresponding form factor. 
A preliminary version of
a paper on the subject is available \cite{Becher Leutwyler}.\\
(b) I then briefly described the work on the large $N_c$ limit in the mesonic
sector, done in 
collaboration with Roland Kaiser. We use a simultaneous expansion of
the effective Lagrangian in powers of derivatives, quark masses and
$1/N_c$ and account for the terms of first non-leading order, as well
as for the one loop graphs -- although these only occur at next-to-next-to 
leading order, they are not irrelevant numerically on account of
chiral logarithms. I drew attention to related work by Feldmann, Kroll
and Stech \cite{Feldmann}. We are currently grinding out the numerical
implications for the various observables of interest. Some of the work is 
described in
ref.~\cite{Kaiser Leutwyler}. A more extensive report is under way.

\newabstract

\begin{center}
  {\large\bf Baryon $\chi$PT in Manifestly Lorentz Invariant Form\footnote{see
  also H. Leutwylers report in these mini-proceedings}}\\[0.5cm]
  {\bf T. Becher} and H. Leutwyler\\[0.3cm]
  Universit\"at Bern, Institut f\"ur theoretische Physik,\\
  Sidlerstr.~5, CH-3012 Bern, Switzerland\\[0.3cm]
\end{center}
The effective theory describing the interaction of pions with a single nucleon
can be formulated in manifestly Lorentz invariant form \cite{GSS}, but it
is not a trivial matter to keep track of the chiral order of graphs containing
loops within that framework: The chiral expansion of the loop graphs in
general starts at the same order as the corresponding tree graphs, so that the
renormalization of the divergences also requires a tuning of those effective
coupling constants that occur at lower order.

Most of the recent calculations avoid this complication by expanding the
baryon kinematics around the nonrelativistic limit. This method is referred to
as heavy baryon chiral perturbation theory \cite{JMBKKM}. It keeps
track of the chiral power counting at every step of the calculation at the
price of manifest Lorentz invariance, but suffers from a deficiency: The
corresponding perturbation series fails to converge in part of the low energy
region.  The problem is generated by a set of higher order graphs involving
insertions in nucleon lines -- this sum diverges.  The problem does not occur
in the relativistic formulation of the effective theory.

The purpose of this talk was to present a method \cite{BL} that exploits the
advantages of the two techniques and avoids their disadvantages. We showed
that the infrared singularities of the one-loop graphs occurring in B$\chi$PT
can be extracted in a relativistically invariant fashion and that this result
can be used to set up a renormalization scheme that preserves both Lorentz
invariance and chiral power counting. The method we have presented follows
the approach of Tang and Ellis \cite{ET}, but we do not expand the infrared
singular part in a chiral series, because that expansion does not always
converge.

\newabstract

\begin{center}
{\large\bf Chiral U(3){\boldmath$\times$\unboldmath}U(3) at
  large {\boldmath$N_{\!c}$\unboldmath}\,: the
  {\boldmath$\eta\!-\!\eta'$\unboldmath} 
  system }\\[0.5cm]
{\bf R. Kaiser} and H. Leutwyler\\[0.3cm]
Institute for Theoretical Physics, University of Bern,\\
Sidlerstr. 5, CH-3012 Bern, Switzerland\\[0.3cm]
\end{center}

In the large $N_{\!c}$ limit the variables required to analyze the low energy
structure of QCD in the framework of an effective field theory necessarily
include the degrees of freedom of the $\eta'$. In a previous analysis
\cite{KL1} we 
demonstrated that the effective Lagrangian prescription of the pseudoscalar
nonet, pions, 
kaons, $\eta$ and $\eta'$, yields results consistent with nature. The
calculation relies on a simultaneous expansion in
powers of momenta, quark masses and $1/N_{\!c}$ which is truncated at first
non-leading 
order. In particular we were able to calculate the decay constants of the
$\eta$ and the $\eta'$ using experimental data  on the decay rates of the
neutral mesons into two photons. The main effect generated
by the corrections to the well known leading order results concerns
$\eta-\eta'$ mixing: at this order of the low energy expansion we need to
distinguish 
two mixing angles. 

The purpose of the present talk is to report on the
results obtained when the above analysis is carried 
further on to systematically include the non-leading corrections to the decay
amplitudes \cite{KL2} (In ref. \cite{KL1} we disregarded the effect of SU(3)
breaking in the 
electromagnetic  
decay rates). Furthermore we include the effects of the one loop graphs:
although algebraically these contributions are suppressed by
one power of $1/N_{\!c}$, some of these are enhanced by logarithmic
factors. The 
points of main interest are the following: (a) is the hypothesis 
that the $1/N_{\!c}$ 
corrections are small valid for the world we live in and (b) what happens to
the low energy theorem for the difference between the two mixing angles if we 
include the one loop graphs. Unfortunately, the analysis now involves a larger
number of unknown low energy constants, so that the numerical analysis yields
ranges for the observables rather than definite values.

\newabstract

\begin{center}
{\large\bf Derivation of the Chiral Lagrangian from the
Instanton Vacuum}\\
[0.5cm]
Dmitri Diakonov\\
[0.3cm]
NORDITA, Blegdamsvej 17, 2100 Copenhagen \O, Denmark \\
[0.3cm]
\end{center}

The QCD vacuum is populated by strong fluctuations of the gluon field
carrying topological charge; these fluctuations are called instantons.
A theory of the QCD instanton vacuum has been suggested in the 80's,
based on the Feynman variational principle \cite{DP1}. In the last
years it has been strongly supported by direct lattice
simulations \cite{Negele}.

Instantons lead to a theoretically beautiful and phenomenologically
realistic microscopic mechanism of chiral symmetry breaking (for a
review see \cite{D1}), which has been also recently checked directly 
on the lattice \cite{Negele}. Chiral symmetry breaking manifests itself 
in quarks acquiring a momentum-dependent dynamical mass, sometimes 
called the constituent mass, and in the appearance of pseudo-Goldstone 
pions interacting with the dynamically massive quarks.  The strength and 
the formafactors of pion-quark interactions are fixed unambiguously by 
the basic parameters of the instanton medium, and agree nicely with
phenomenology \cite{D2}.

Integrating out quarks, one gets the chiral lagrangian, whose
low-momentum expansion and the asymptotic form have been
investigated \cite{D2}.

\newabstract

\begin{center}
{\large\bf Theoretical Study of Chiral Symmetry Breaking in QCD: Recent
Development}\\[0.5cm]
Jan Stern\\[0.3cm]
Groupe de Physique Theorique, IPN,\\
91406 Orsay, France\\[0.3cm]
\end{center}
If forthcoming experiments confirmed the actual value of $\pi\pi$
scattering length $a^0_0=0.26 \pm 0.05$ with a smaller error bar, we would
have to conclude that the condensate $<\bar q q>$ is considerably smaller than
one usually believes.
I have presented few theoretical speculations inspired
by this possibility: 1) Chiral symmetry breaking is due to the accumulation
of small eigenvalues of the Euclidean Dirac operator in the limit of a large
volume V. Since the fermion determinant reduces the weight of small
eigenvalues, order parameters $F^2_\pi $ and $<\bar q q>$ are suppressed for
large $N_f / N_c$  and the symmetry will be restored for $N_f/N_c >n_0 $ ,
(likely, $n_0=3.25$). The condensate could disappear more 
rapidly than $F^2_\pi$,
since it is merely sensitive to the smallest eigenvalues behaving as $1/V$.
It is even conceivable that for $n_1<N_f/N_c<n_0$, there is a phase in which
$<\bar q q> =0$ but $F_\pi \ne 0$, implying symmetry breaking and
the existence of Goldstone bosons\cite{Stern1}.     
Close to the critical point $n_1$, it
would then be natural to expect a tiny quark condensate. 2)    
Kogan, Kovner and Shifman
\cite{Kogan} have pointed out that due to Weingarten's inequality, the bare,
cutoff dependent condensate cannot vanish unless $F_\pi =0$. This, however,
does not exclude vanishing of the renormalized condensate (assuming QCD-like
sign of its anomalous dimension) and the existence of the critical point
$n_1<n_0$ remains an open possibility. 3) Below but close to $n_1$ one
should  expect important Zweig rule
violation and huge corrections to large $N_c$ predictions \cite{Stern2}.
4) The critical point $n_1$ could be seen through the volume dependence of
Leutwyler-Smilga sum rules for Dirac eigenvalues \cite{Descotes}.
The latter could be
investigated numerically,  diagonalizing the discretized square of
the continuum massless Dirac
operator, thereby avoiding the well known difficulties with massless fermions
on the lattice .

\newabstract

\begin{center}
{\large\bf NUMERICAL SOLUTIONS OF ROY EQUATIONS}\\[0.5cm]
Gilberto Colangelo \\
Institut f\"ur Theoretische Physik, Universit\"at Z\"urich \\
Winterthurerstr. 190, CH--8057 Z\"urich\\[0.3cm]
\end{center}

\noindent
Roy equations \cite{Roy} are a system of coupled integral equations for the
partial wave amplitudes of $\pi \pi$ scattering that incorporate the
properties of analyticity, unitarity and crossing symmetry.  Besides the
dispersive integrals containing the Roy kernels and the partial wave
amplitudes, the equations contain polynomial terms which depend on two
constants only, i.e. the two $S$-wave scattering lengths. Solutions of Roy
equations depend on the input values of these two constants.

I have reported about recent work \cite{roy_coll} in solving numerically
the Roy equations, describing both the method and the results. Our aim is
to extract information on the low-energy behaviour of the $\pi \pi$
scattering partial wave amplitudes. For what concerns the high energy part,
we use all the available information (coming both from experimental data
and, where these are not available, from theoretical modelling) as input in
the Roy equations, and solve them numerically only in the low-energy
region.  Since unitarity, analyticity and crossing do not constrain in any
way the two $S$-wave scattering lengths on which the solution depends, we
need additional information on those. This information comes from chiral
symmetry, and to provide it we match the chiral representation of the
amplitude (now available at the two loop level \cite{BCEGS}) to the
dispersive one, at low energy.

I have detailed how this matching is done, and have shown that the outcome
of this combination of two different theoretical tools, together with the
experimental information available on this process, is a very precise
representation for the $\pi \pi$ scattering amplitude at low energy. New
$K_{l4}$ data on the $\pi \pi$ phase shifts near threshold, expected in the
near future from the E865 collaboration at BNL and from KLOE at DA$\Phi$NE
(the Frascati $\Phi$ factory), will test the validity of this
representation with high accuracy. A different experimental test will also
come from the measurement of the pionic atoms lifetime made by the DIRAC
collaboration: this measurement will give direct access to the difference
of the two $S$--wave scattering lengths.

\newabstract
\begin{center}
{\large\bf One-channel Roy equation revisited}\\[0.5cm]
{\bf J. Gasser}$^1$ and {G. Wanders}$^2$\\[0.3cm]
$^1$ Institut f\"ur  Theoretische Physik, Universit\"at  Bern,\\
Sidlerstrasse 5, CH--3012 Bern, Switzerland \\[.3cm]
 $^2$ Institut de Physique Th\'eorique, Universit\'e  de Lausanne,\\
 CH--1015 Lausanne, Switzerland\\[.3cm]
\end{center}
 The Roy
equation \cite{roy} in the single channel case
amounts to a nonlinear, singular integral equation for the phase 
shift $\delta$ in the
low--energy  region,
\begin{eqnarray}\label{eqone}
\frac{1}{2\sigma(s)}\sin{[2\delta(s)]}&=&a+\frac{(s-4)}{\pi}
P\hspace{-.38cm}\int_4^\infty
\frac{dx}{x-4}\frac{\omega(x)}{x-s}\;,\;\nonumber\\
\omega(x)&=&\left\{\begin{array}{cl}
\sigma(x)^{-1}\sin^2[\delta(x)] &; \;  4 \leq x \leq s_0\\
A(x) & ; \;x \geq s_0 \;,
\end{array}\right..
\end{eqnarray}
Here, $\sigma(s)=(1-4/s)^{1/2}$ is the phase space factor, and $A(x)$
denotes the imaginary part of the partial wave 
above the 
matching point $s_0$. The integral is a principal value one. The
 problem consists in solving (\ref{eqone}) in the 
 interval $[4,s_0]$ at given  scattering length $a$ and 
 given imaginary part $A$.
Investigating the infinitesimal neighborhood of
a solution, the
 following  proposition  was established  some time ago \cite{pw}:

 {\it 
 Let $\delta$ be a solution of
equation (\ref{eqone}) with input $(a,A)$. It is an isolated solution if
$-\frac{\pi}{2} < \delta(s_0)<\frac{\pi}{2}$. If $\delta(s_0)>
\frac{\pi}{2}$, the infinitesimal neighborhood of $\delta$ is an
$m$--parameter family of solutions $\delta'$ with
$\delta'(s_0)=\delta(s_0)$, where $m$ is the integral part of
$2\delta(s_0)/\pi$ for $\delta(s_0)>\pi/2$, and zero otherwise.}

In \cite{gw}, we recall the derivation of this proposition
  -- in particular, we detail  its  connection with the
 homogeneous Hilbert
 problem on a finite interval. In 
 addition, we construct explicit 
 expressions for amplitudes that solve the full, nonlinear  Roy
equation (\ref{eqone}) exactly.
These amplitudes contain free parameters that render  the 
non--uniqueness of
 the solution manifest. The amplitudes
 develop, however,  in general an unphysical  singular behaviour at 
 the matching point $s_0$.
  This singularity disappears and uniqueness
is achieved if one uses analyticity properties of the amplitudes that
are not encoded in the Roy equation.

\newabstract

\begin{center}
{\large\bf How do the uncertainties on the input affect the solution of the
Roy equations?}\\[0.5cm]
G\'erard Wanders\\[0.3cm]
Institut de physique th\'eorique, Universit\'e de Lausanne,\\
CH-1015 Lausanne, Switzerland\\
E-mail: Gerard.Wanders@ipt.unil.ch\\[0.3cm]
\end{center}

An update of previous work~\cite{Pom1},\cite{Epe2} on the
possible non-uniqueness of
the solution of $S$- and $P$-wave Roy equations~\cite{Roy3} and its
sensitivity to the uncertainties in the input is under way. The Roy equations
determine the $S$- and $P$-wave phase shifts in a low-energy interval
$[2M_\pi,E_0]$. The $S$-wave scattering lengths and absorptive parts above
$E_0$ are part of the input.

The Roy equations are used nowadays for the extrapolation of the available
data down to threshold. Colangelo~et~al.~\cite{Col5} take
$E_0=800$~MeV, a value for which the solution is unique [there is a continuous
family of solutions if $E_0>860$~MeV]. The result of a variation of the input
has been derived in~\cite{Epe2} for $E_0=1.1$~GeV and this is redone now for
$E_0=800$~meV. The linear response to small variations of the scattering
lengths has been determined using a model with $a_0^0=0.2$ and $a_0^2=-0.041$.
The effect is strongest on the $P$-wave. The phase shift $\delta_1^1$ is
changed into $\delta_1^1+\Delta_1^1$ and the ratio $\Delta_1^1/\delta_1^1$ is
of the same order of magnitude as $\delta a_0^0/a_0^0$ or $\delta a_0^2/a_0^2$
over the whole interval $[2M_\pi,E_0]$ when $a_0^0$ and $a_0^2$ are varied
separately. According to the previous talk~\cite{Gas6}, $\Delta_1^1/\delta_1^1$
exhibits a cusp at $E_0$. This cusp is extremely sharp and plays a role in the
numerical analysis~\cite{?7}. In conformity with a universal curve
in the $(a_0^0,a_0^2)$-plane, $\Delta_1^1\sim(\delta a_0^0-4.9\delta a_0^2)$
over $[2M_\pi,E_0]$. The effects on the $S$-waves are less spectacular and the
cusps are practically invisible.

\newabstract

\begin{center}
{\large\bf Properties of a possible scalar nonet}\\[0.5cm]
Deirdre Black$^1$, Amir Fariborz$^1$, Francesco Sannino$^2$\\ and {\bf
Joseph Schechter}$^1$\\[0.3cm]
$^1$Physics Dept, Syracuse University,\\ Syracuse NY 13244-1130,
USA\\[0.3cm]
$^2$Physics Dept, Yale University,\\
New Haven CT 06520-8120\\[0.3cm]
\end{center}

It was found that a light sigma-type meson and a light kappa-type meson
are needed to preserve unitarity in, respectively, models of $\pi\pi$
\cite{1} and $\pi K$\cite{2} scattering. These models are based on a
chiral Lagrangian and yield simple approximate amplitudes which satisfy
both crossing symmetry and the unitarity bounds. Together with the well
established $f_0(980)$ and $a_0(980)$ mesons these would fill up a low
lying nonet. We investigate the "family" relationship of the members of
this postulated nonet. We start by considering Okubo's original
formulation\cite{3} of "ideal mixing". It is noted that the original mass
sum rules possess another solution which has a natural interpretation
describing a meson nonet made from a dual quark and a dual anti-quark.
This has the same structure as a model for the scalars proposed\cite{4}
by Jaffe in the context of the MIT bag model. However, our masses do
not exactly satisfy the alternate ideal mixing sum rule. For this reason,
and also to obtain the experimental pattern of decay modes, we introduce
additional terms which break ideal mixing. Then two different solutions
giving correct masses but characterized by different scalar mixing
angles emerge. The solution which yields decay widths agreeing with
experiment has a mixing angle about $-17^o$ while the other has
a mixing angle about $-90^o$. For comparison, ideal mixing for
a quark anti-quark nonet is $\pm 90^o$ while ideal mixing for
a dual quark, dual anti-quark nonet is $0^o$. Hence the dual
picture is favored. A more detailed description of this work is given
in the report\cite{5}.

\newabstract

\begin{center}
{\large\bf ChPT Phenomenology in the Large--$N_C$ Limit}\\[0.5cm]
{\bf A. Pich}\\[0.3cm]
Department of Theoretical Physics, IFIC, Univ. Valencia -- CSIC\\
Dr. Moliner 50, 46100 Burjassot (Valencia), Spain.\\[0.3cm]
\end{center}

Chiral symmetry provides strong low--energy constraints on the dynamics of
the Goldstone bosons. However, we need additional input to analyze
physics at higher energy scales. The following two examples show that
very useful information can be obtained from the large--$N_C$ limit of QCD:

\begin{itemize}

\item Using our present knowledge on effective hadronic theories, 
short--distance QCD information, the $1/N_C$ expansion, analyticity 
and unitarity, we derive
an expression for the pion form factor \cite{GP:97} in
terms of $m_\pi$, $M_\rho$ and $f_\pi$. This parameter--free
prediction provides a surprisingly good description of the experimental
data up to energies of the order of 1~GeV.
A similar analysis of the $K\pi$ scalar form factor, needed for the 
determination of the strange quark mass, is in progress.

\item The dispersive two--photon contribution to the $K_L\to\mu^+\mu^-$
decay amplitude is analyzed using chiral perturbation theory techniques 
and large--$N_C$ considerations. A consistent description \cite{GP:98}
of the decays
$\pi^0\to e^+e^-$, $\eta\to\mu^+\mu^-$ and $K_L\to\mu^+\mu^-$ is obtained.
Moreover, the present data allow us to derive a useful constraint on the
short--distance $K_L\to\mu^+\mu^-$ amplitude, which could be improved by
more precise measurements of the $\eta\to\mu^+\mu^-$ and $K_L\to\mu^+\mu^-$
branching ratios. This offers a new possibility for testing the
flavour--mixing structure of the Standard Model.
As a by--product one predicts 
$B(\eta\to e^+e^-) = (5.8\pm 0.2) \times 10^{-9}$ and
$B(K_L\to e^+e^-) = (9.0\pm 0.4) \times 10^{-12}$.

\end{itemize}

\newabstract

\begin{center}
{\large\bf Gauge-invariant Effective Field Theory \\ 
  for a Heavy Higgs Boson}\\[0.4cm] 
{\bf Andreas Nyf\/feler}$^1$ and Andreas Schenk$^2$\\[0.2cm]
$^1$DESY, Platanenallee 6, D-15738 Zeuthen, Germany; 
nyf\/feler@ifh.de \\[0.1cm]
$^2$Hambacher Stra\ss e 14, D-64625 Bensheim-Gronau, Germany.\\[0.3cm] 
\end{center}

The method of effective field theory has repeatedly been used to
analyze the symmetry breaking sector of the Standard Model and to
parametrize effects of new physics.  Comparing the theoretical
predictions for the low energy constants in the corresponding
effective Lagrangian for different models with experimental
constraints might help to distinguish between different underlying
theories before direct effects become visible.  At low energies the
Standard Model with a heavy Higgs boson in the spontaneously broken
phase, which serves as a reference point for a strongly interacting
symmetry breaking sector, can adequately be described by such an
effective field theory.  Moreover, the low energy constants can
explicitly be evaluated using perturbative methods by matching the
full and the effective theory at low energies.  Several groups have
performed such a calculation for a heavy Higgs boson in recent
years~\cite{HM_EM_DGK}. In gauge theories there are, however, some
subtleties involved when this matching is performed with
gauge-dependent Green's functions. We therefore proposed a manifestly
gauge-invariant approach in Ref.~\cite{Abelian_Higgs} which deals only
with gauge-invariant Green's functions.

In this talk we presented the extension of our method to the
non-abelian case. Using a generating functional of gauge-invariant
Green's functions for the bosonic sector of the Standard Model which
was discussed recently~\cite{SM_gaugeinv}, we evaluated the
effective Lagrangian for a heavy Higgs boson at the one-loop level and
at order $p^4$ in the low-energy expansion by matching corresponding
gauge-invariant Green's functions at low energies.  A detailed
description of our calculation which preserved gauge invariance
throughout the matching procedure and a comparison of our results with
those obtained by the other groups~\cite{HM_EM_DGK} can be found in
Ref.~\cite{Heavy_Higgs_gaugeinv}.

\newabstract

\begin{center}
{\large\bf The Goldberger-Treiman Discrepancy in SU(3)}\\[0.5cm]
{\bf J. L. Goity}$^1$, R. Lewis$^2$, M. Schvellinger$^3$, and L. Zhang$^1$\\[0.3cm]
{\it $^1$Dept.   Phys., Hampton University,
Hampton, VA 23668, USA,\\
and Jefferson Lab, Newport News, VA 23606, USA. \\
$^2$Dept. of Phys., University of Regina, Regina, Canada.\\
$^3$Dept. of Phys., Universidad Nac. de La Plata, La Plata, Argentina.\\[0.3cm]}
\end{center}

We  studied the Goldberger-Treiman discrepancy (GTD)\cite{GTD} in the
baryon octet 
in the framework of  HBChPT. We confirm  the
previous conclusion\cite{Gasser},\cite{HUGS} that at
leading order (order $p^2$)
the discrepancy is entirely given by contact terms in the baryon effective 
Lagrangian of order $p^3$, and demonstrate that the subleading corrections
are of order $p^4$.
At leading order and in the isospin symmetry limit there are only two terms
that affect the GTD, namely,
\begin{eqnarray}
{\cal {L}}^{(3)}_{GTD} &=& F_{19} {\rm Tr} ({\bar{B}} S_v^\mu
[\nabla_\mu\chi_{-}, \, B]) \nonumber \\
&+& D_{19} {\rm Tr} ({\bar{B}} S_v^\mu \{\nabla_\mu\chi_{-}, \, B\}),
 \nonumber  
\label{L3}
\end{eqnarray} where the notation is standard.
We analyze the discrepancies that can be determined from the available 
meson-nucleon
couplings, namely  $\pi{N}N$, $KN\Lambda$ and $KN\Sigma$, and conclude that 
only 
the smaller values of the
$g_{\pi NN}$ coupling lead to a consistent picture where the 
$p^4$ corrections to the discrepancies would have natural size. We also
check that with the smaller values of $g_{\pi NN}$ the
Dashen-Weinstein relation is well satisfied,
while it is badly violated for the larger values.

\newabstract
\begin{center}
{\large\bf Tau decays and Generalized $\chi$PT}\\[0.5cm]
Luca Girlanda\\[0.3cm]
Groupe de Physique Th\'eorique, Institut de Physique Nucl\'eaire,\\
F-91406 Orsay, France.\\[0.3cm]
\end{center}
The hadronic matrix element  of the decays $\tau \rightarrow 3 \pi \nu_{\tau}$
 contains a large spin-1 part, dominated by the $a_1$ resonance, and a small
 spin-0 part, which is proportional to the divergence of the axial current
 and then to the light quark mass. The latter, involving the explicit chiral
 symmetry breaking sector of the theory, is expected to be very sensitive to
 the size of the quark anti-quark condensate $\langle \bar q q \rangle$.
 In order to investigate this sensitivity we use the generalized version of
 $\chi$PT \cite{gchpt}.

The $SU(2) \times SU(2)$ generalized chiral lagrangian is constructed and
 renormalized at one-loop level.
 We then compute the nine structure functions (see Ref.~\cite{km}) for both
 the charge modes $\tau \rightarrow \pi^- \pi^- \pi^+ \nu_{\tau}$ and  $\tau
 \rightarrow \pi^0 \pi^0 \pi^- \nu_{\tau}$.
% The results are expressed in terms of the parameters $\alpha$ and $\beta$
% of the elastic  $\pi\pi$ amplitude, introduced in Ref.~\cite{pipi}.
% They are directly correlated to the size of $\langle \bar q q \rangle$ (see
% Ref.~\cite{ioparigi}).
 In the limit of a large condensate we recover the results of standard
 $\chi$PT, found in Ref.~\cite{gilberto}.

The spin-1 contribution is kinematically suppressed, at threshold, leading
 to sizeable azimuthal asymmetries which depend strongly on the size of
 $\langle \bar q q \rangle$, up to rather high values of the hadronic
 invariant mass $Q^2$. The integrated left-right asymmetry for the
 all-charged mode,
 for instance, varies from $(28 \pm 4) \%$ up to $(60 \pm 6) \%$ at $Q^2 = 0.35\,\,
 \mbox{GeV}^2$ if the condensate decreases from its standard value down to
 zero\footnote{
 These numbers correspond to taus produced at rest, and thus they are
 relevant only for the $\tau$-charm factories. Minor modifications in the
 analysis are needed to account for ultrarelativistic taus, currently
 produced in the accelarators.}.

Unfortunately the branching ratio in this kinematical region is quite
 small: the largest statistics up to now has been collected by the CLEO
 Collaboration, with about $10^7$ $\tau$-pairs \cite{perl},
  corresponding to about 100 events from threshold to $Q^2 = 0.35\,\,
 \mbox{GeV}^2$.

\newabstract

\begin{center}
{\large \textbf{Automatized ChPT Calculations}}\\[0.5cm]
Frederik Orellana$^{1}$\\[0.3cm]
$^{1}$Inst. Theor. Phys., University of Z\"{u}rich,\\[0pt]
Winterthurerstr. 190, 8057 Z\"{u}rich, Switzerland\\[0.3cm]
\end{center}

In recent years much effort has been put into automatizing Feynman diagram
calculations in the Standard Model and in the Minimal Super Symmetric
Standard Model (see \cite{AIHEP}, \cite{TEFE}, \cite{HAPE} and references
therein). The reason is the increasing complexity of calculations due to
more loops, more couplings and intermediate particles and more particles in
the final state. These complications are motivated by the increasing energy
and precision of the experiments.

It is argued that despite the different nature of ChPT \cite{GL1}, it is
interesting and feasible also here to apply some automatization. Some of the
different features of ChPT are due to the fact that it is a
non-renormalizable effective theory with an expansion in the momentum,
multi-leg vertices and new couplings at each order in the expansion,
expressed in a rather compact notation.

An overview of existing computer programs is given. A framework is proposed
for automatizing ChPT calculations \cite{FA}, \cite{FC} and a new program is
presented \cite{PHI}.

\newabstract

\begin{center}
{\large\bf From Chiral Random Matrix Theory \\
to Chiral Perturbation Theory}\\[0.1cm]
J.C. Osborn, {\bf D. Toublan} and J.J.M. Verbaarschot\\[0.1cm]
Department of Physics and Astronomy, \\
SUNY, Stony Brook, New York 11794.\\[0.1cm]
\end{center}

\begin{small}
We study the spectrum of the QCD Dirac operator by means of
the valence quark mass dependence of the chiral condensate in 
partially quenched Chiral Perturbation Theory in the 
supersymmetric formulation of Bernard and Golterman \cite{pqChPT}. 
We consider valence quark masses both in the ergodic domain 
($m_v \ll E_c$) and the diffusive domain ($m_v \gg E_c$). 
These domains are separated by a mass scale
$E_c \sim F^2/\Sigma_0 L^2$ (with $F$ the pion decay constant, $\Sigma_0$ the 
chiral condensate and $L$ the size of the box) \cite{OV}. We perform
a finite volume analysis of partially quenched Chiral Perturbation Theory 
along the same lines as the one done in Chiral Perturbation Theory \cite{GLS}.

In the ergodic domain the effective super-Lagrangian
reproduces the microscopic spectral density of chiral Random 
Matrix Theory \cite{Cambridge},\cite{OTV}.

In the diffusive domain we extend the results for the slope of the
Dirac spectrum first obtained by Smilga and Stern \cite{SS}.
We find that the spectral density diverges logarithmically for nonzero (sea) 
quark masses. We study the transition
between the ergodic and the diffusive domain and identify a range where
chiral Random Matrix Theory and partially quenched Chiral Perturbation 
Theory coincide \cite{OTV}. 

We also point out some interesting analogies with mesoscopic disordered 
systems and with chaotic ones \cite{Bohigas}.
\\[-1cm]

\end{small}

\end{document}